\documentclass[a4paper]{cas-sc}

\usepackage[authoryear,longnamesfirst]{natbib}

\newcommand{\be}{\begin{equation}}
\newcommand{\ee}{\end{equation}}
\usepackage{amsmath,amssymb}
\usepackage{mathtools}
\usepackage{amsthm}
\usepackage{algorithm}
\usepackage{algpseudocode}
\algrenewcommand\algorithmicrequire{\textbf{Inputs:}}
\algrenewcommand\algorithmicensure{\textbf{Outputs:}}

\newtheorem{lemma}{Lemma}
\newtheorem{assumption}{Assumption}
\newtheorem{theorem}{Theorem}
\newtheorem{remark}{Remark}
\newtheorem{corollary}{Corollary}

\usepackage{multirow}
\def\tsc#1{\csdef{#1}{\textsc{\lowercase{#1}}\xspace}}
\tsc{WGM}
\tsc{QE}
\tsc{EP}
\tsc{PMS}
\tsc{BEC}
\tsc{DE}

\makeatletter
\def\ps@first{%
  \let\@oddhead\@empty
  \let\@evenhead\@empty
  \def\@oddfoot{\hfil\thepage\hfil}%
  \let\@evenfoot\@oddfoot
}
\def\ps@plain{%
  \let\@oddhead\@empty
  \let\@evenhead\@empty
  \def\@oddfoot{\hfil\thepage\hfil}%
  \let\@evenfoot\@oddfoot
}
\makeatother

\begin{document}
\let\WriteBookmarks\relax
\def\floatpagepagefraction{1}
\def\textpagefraction{.001}


\shorttitle{Conformalized Robust Principal Component Analysis}

\shortauthors{Anonymous}

\title [mode = title]{Conformalized Robust Principal Component Analysis}                     

\author[1,2]{Liangliang Yuan}

\ead{llyuan_dlut@163.com}

\affiliation[1]{organization={School of Mathematics and Statistics, Nanjing University of Science and Technology}, country={China}}

\author[2]{Lei Wang}[style=chinese]
\ead{lwangstat@nankai.edu.cn}
\affiliation[2]{organization={School of Statistics and Data Science, KLMDASR, LEBPS and LPMC, Nankai University}, country={China}}

\author[3]{Quan Kong}
\affiliation[3]{organization={School of Cyber Science and Engineering, Nanjing University of Science and Technology}, country={China}}

\author[4]{Liuhua Peng}
\affiliation[4]{organization={School of Mathematics and Statistics, University of Melbourne}, country={Australia}}




\begin{abstract}
Robust principal component analysis (RPCA) is a widely used technique for recovering low-rank structure from matrices with missing entries and sparse, possibly large-magnitude corruptions. Although numerous algorithms achieve accurate point estimation, they offer little guidance on the uncertainty of recovered entries, limiting their reliability in practice. In this paper,  we propose conformal prediction-RPCA (CP-RPCA), a practical and distribution-free framework for uncertainty quantification in robust matrix recovery. Our proposed method supports both split and full conformal implementations and incorporates weighted calibration to handle heterogeneous observation probabilities. We provide theoretical guarantees for finite-sample coverage and demonstrate through extensive simulations that CP-RPCA delivers reliable uncertainty quantification under severe outliers, missing data and model misspecification. Empirical results show that CP-RPCA can produce informative intervals and remain competitive in efficiency when the RPCA model is well specified, making it a scalable and robust tool for uncertainty-aware matrix analysis.
\end{abstract}




\maketitle

\section{Introduction}

Robust principal component analysis (RPCA) seeks to decompose an observed data matrix into a low-rank component capturing latent structure and a sparse component accounting for outliers or gross corruptions. Compared with classical PCA, RPCA explicitly models structured noise and large anomalies, making it well suited for high-dimensional data affected by heavy-tailed noise, occlusion, or adversarial contamination \citep{candes2011robust, bertsimas2023sparse}. Owing to its robustness and interpretability, RPCA has become a core tool in a wide range of applications, including video background–foreground separation \citep{yi2016fast, zhang2018unified, xu2024lere}, robust face recognition \citep{wright2009robust}, network denoising \citep{shao2023distribution}, anomaly detection \citep{chen2011integrating}, collaborative filtering \citep{xu2010robust}, bioinformatics \citep{myers2025chronological} and system identification \citep{vaswani2018robust, menon2019structured}.

Methodologically, the existing RPCA approaches can be broadly divided into \textit{convex relaxation} methods and \textit{nonconvex optimization} methods. Convex approaches replace the rank function and the $\ell_0$ penalty by the nuclear norm and the $\ell_1$ norm, respectively, and achieve exact recovery under incoherence and sparsity conditions \citep{candes2011robust, chandrasekaran2011rank}. Subsequent work improved statistical guarantees and computational scalability under noisy observations and higher corruption levels \citep{agarwal2012noisy, chen2013low, klopp2017robust}. Despite their strong theoretical guarantees, convex methods typically require repeated singular value decompositions, which can be computationally prohibitive in large-scale settings. To address these limitations, nonconvex methods based on low-rank matrix factorization have been actively developed. \cite{chen2015fast} proposed a projected gradient descent algorithm for RPCA with provable fast convergence under suitable initialization, while \cite{wang2017unified} established a unified recovery theory for nonconvex low-rank estimation in rectangular matrices. Building on these foundations, \cite{yi2016fast} and \cite{zhang2018unified} introduced truncation- and thresholding-based algorithms for RPCA, which achieve improved sample and computational efficiency while maintaining strong robustness to sparse outliers.

More recently, \cite{chen2021bridging} provided a unifying theoretical framework connecting convex and nonconvex RPCA methods. They showed that, with appropriate initialization and regularization, nonconvex approaches can achieve the same statistical recovery guarantees as nuclear-norm minimization, while substantially reducing computational cost. This result clarifies the relationship between the two paradigms and offers theoretical guidance for designing efficient and statistically principled RPCA algorithms.

\subsection{Related work}

Despite substantial progress in algorithmic development for low-rank matrix recovery, uncertainty quantification for the resulting estimators remains far less understood. Closely related challenges also arise in high-dimensional structured data problems such as matrix and tensor completion, principal component analysis and  network analysis. Existing work on uncertainty quantification in these settings broadly falls into two categories. The first category comprises \textit{model-based} inference methods, which rely on explicit low-rank structural assumptions and specific noise distributions. In matrix completion, \cite{chen2019inference} studied debiased estimators for both convex and nonconvex methods and established asymptotic normality and entrywise confidence intervals under incoherence and Gaussian noise assumptions. These ideas have been extended to tensor settings \citep{cai2022uncertainty, xu2025statistical} and to principal component analysis with heteroskedastic noise and missing data \citep{yan2024inference}. Related works further consider heterogeneous or heavy-tailed noise \citep{farias2022uncertainty}, inference for linear functionals of missing entries \citep{xia2021statistical} and  statistical inference for noisy 1-bit matrix completion \citep{chen2023statistical}. While theoretically powerful, these approaches typically require stringent assumptions on the low-rank structure, noise distribution and  observation mechanism, and their validity can deteriorate when such assumptions are violated. To address this limitation, a second line of research develops \textit{distribution-free} or \textit{model-agnostic} uncertainty quantification methods, often based on conformal inference. Conformal prediction provides finite-sample coverage guarantees without parametric assumptions and has recently been adapted to structured matrix problems \citep{zhou2025conformal}. In particular, \cite{gui2023conformalized} proposed conformalized matrix completion, delivering valid prediction intervals for missing entries under arbitrary model misspecification. Subsequent work incorporated additional structural information to construct joint confidence regions \citep{liang2024structured} and  extended conformal inference to tensor completion via Riemannian optimization \citep{sun2025conformalized}. Beyond matrix and tensor settings, \cite{shao2023distribution} developed one of the first distribution-free uncertainty quantification frameworks for network data with missing entries, introducing coverage notions tailored to dependent graph structures. Together, these works highlight the potential of conformal and model-agnostic approaches for uncertainty quantification in high-dimensional structured data.

Notwithstanding the rapid development of distribution-free uncertainty quantification methods for matrix, tensor and  network data, their extension to RPCA remains largely unexplored. RPCA is a fundamental tool for analyzing high-dimensional data contaminated by sparse but potentially large corruptions, where reliable statistical inference requires more than accurate point estimation. In applications such as video background modeling, financial monitoring and  bioinformatics, uncertainty quantification for the recovered low-rank component is essential for assessing estimation stability, identifying high-risk regions and  supporting downstream analysis. These considerations highlight the need for principled, distribution-free inference methods tailored to RPCA. Our proposed method builds on conformal prediction (CP), a distribution-free framework for uncertainty quantification introduced by \cite{vovk2005algorithmic}. Under data exchangeability, CP yields finite-sample valid prediction intervals or confidence sets for arbitrary data distributions and black-box predictors, without relying on parametric assumptions \citep{angelopoulos2024theoretical}. Although the full conformal procedure enjoys strong theoretical guarantees, it is computationally expensive. To address this issue, \cite{lei2018distribution} proposed split CP, \cite{vovk2015cross} developed cross-conformal inference, and \cite{barber2021predictive} introduced jackknife and jackknife+ methods, for substantially improving scalability and preserving finite-sample validity. To handle departures from exchangeability, such as covariate shift, \cite{tibshirani2019conformal} proposed weighted CP, establishing coverage guarantees under weighted exchangeability. Building on this idea, \cite{barber2023conformal} studied conformal inference under more general distributional shifts and quantified the resulting coverage loss. For comprehensive treatments of conformal inference and its extensions, see \cite{angelopoulos2021gentle}, \cite{angelopoulos2024theoretical} and  \cite{zhou2025conformal}.

\subsection{Our contributions}

In this paper, we propose a practical and distribution-free CP-RPCA framework for uncertainty quantification in robust matrix recovery. Our main contributions are summarized as follows:

\begin{enumerate}[\textbullet]
\item Our proposed method provides entrywise confidence intervals for the recovered low-rank component with validity independent of the specific estimation algorithm. Different from a direct extension of \cite{gui2023conformalized}, our approach adopts a two-stage procedure to explicitly account for the impact of sparse outliers on uncertainty quantification. Moreover, in contrast to \cite{chen2021bridging}, we do not assume that the support of the sparse component is contained in the observed index set, which is more closely reflects practical data contamination scenarios.
\item Compared with the existing CP-based uncertainty quantification methods for high-dimensional random matrices, we establish both non-asymptotic lower and upper bounds on the coverage probability of the proposed confidence intervals, thereby strengthening the associated theoretical guarantees.
\item We demonstrate the practical utility of CP-RPCA in face recognition under complex environments, where the lengths of confidence intervals are used to automatically identify high-uncertainty regions, guiding finer-grained feature modeling and improving both accuracy and interpretability.
\end{enumerate}

The remainder of this paper is organized as follows. Section 2 presents the conformalized RPCA framework, including the problem setup and two-stage procedures based on split and full CPs. Section 3 develops the theoretical foundations of CP-RPCA, establishing weighted exchangeability and finite-sample coverage guarantees. Section 4 evaluates the finite-sample performance of CP-RPCA through extensive simulation studies. In Section 5, we demonstrate the practical utility of the proposed method via two real-data applications: feature extraction in face images and video background modeling. Section 6 concludes with a summary of the main findings and directions for future research. Supplementary materials contain technical proofs, additional methodological details and  simulation results.

\noindent\textbf{Notation.} 
For any matrix $A \in \mathbb{R}^{d_1 \times d_2}$, let $A_{i,\ast}$ and $A_{\ast,j}$ denote its $i$-th row and $j$-th column, respectively, and let $A_{ij}$ denote the $(i,j)$-th entry. We write $\mathrm{SVD}_r(A)$ for the rank-$r$ singular value decomposition of $A$. The spectral norm and Frobenius norm of $A$ are denoted by $\lVert A \rVert_2$ and $\lVert A \rVert_F$, respectively. The number of nonzero entries of $A$ is denoted by $\lVert A \rVert_0$, the maximum absolute entry by $\lVert A \rVert_{\infty ,\infty}$ and the maximum row $\ell_2$-norm by $\lVert A \rVert _{2,\infty}$. For any $t \in \mathbb{R}$, $\delta_t$ denotes the point mass distribution at $t$. Unless otherwise specified, $\Omega^{\rm{c}}$ denotes the complement of a set $\Omega$ and $|\Omega|$ denotes its cardinality. For two sets $\Omega_1$ and $\Omega_2$, the set difference is defined as $\Omega_1 \setminus \Omega_2=\Omega_1\cap\Omega_2^{\rm{c}}$.

\section{Conformalized RPCA}

\subsection{Problem setup}

As discussed in the classical RPCA literature, we consider a partially observed data model of the form
$$
Y_{ij} = X_{ij}^{\ast} + S_{ij}^{\ast} + E_{ij},~~(i, j)\in \Omega_{\mathrm{obs}},
$$
where $Y=[Y_{ij}]\in \mathbb{R}^{d_1 \times d_2}$ denotes the observed data matrix containing noise, outliers and missing entries,
$X^{\ast}=[X_{ij}^{\ast}] \in \mathbb{R}^{d_1 \times d_2}$ is a low-rank matrix capturing the latent structure, $S^{\ast}=[S_{ij}^{\ast}] \in \mathbb{R}^{d_1 \times d_2}$ is a sparse matrix representing gross corruptions and $E=[E_{ij}] \in \mathbb{R}^{d_1 \times d_2}$ denotes random noise, typically assumed to have i.i.d. (sub-)Gaussian entries. Observations are available only on an index set $\Omega_{\mathrm{obs}}\subseteq [d_1]\times[d_2]$, where $[d]:=\{1,2,\ldots,d\}$. We denote by
$$
\Omega^{\ast}:=\{(i, j)\in [d_1]\times[d_2]: S_{ij}^{\ast}\neq 0\}
$$
the support of the sparse corruption matrix $S^{\ast}$.\footnote[1]{ we do not assume that the support of the sparse component is contained in the observed index set, which better reflects practical scenarios with arbitrary corruptions.} Classical RPCA theory ensures identifiability by imposing incoherence conditions on $X^{\ast}$ and sparsity constraints on $S^{\ast}$; see, for example, \cite{yi2016fast} and \cite{chen2021bridging}. Under these assumptions together with a known or correctly specified rank and homogeneous missingness, existing work primarily focuses on point estimation and establishes error bounds for recovering $X^{\ast}$ and $S^{\ast}$. Unfortunately, these assumptions are often violated in practice and then a fundamental but less studied question is how to quantify the reliability of these point estimates robustly.

To address this issue, in this paper we investigate uncertainty quantification for the low-rank component in RPCA under substantially weaker, distribution-free assumptions. Importantly, we retain the low-rank plus sparse decomposition \citep{sun2025conformalized}, since the low-rank component provides a compact representation for dimension reduction and compression, while the sparse component captures localized outliers and corruptions, enabling robustness. Moreover, motivated by \cite{gui2023conformalized} and noticing that the low-rank component often corresponds to a stable background structure in applications (Section~\ref{sec5.2}), we treat it as a fixed but unknown matrix when performing uncertainty quantification. Specifically, we only assume  that the observed matrix admits a decomposition
$$
Y=X+S^{\ast},
$$
where $X$ is approximately low rank and $S^{\ast}$ is sparse. All randomness in our framework arises from the observation process and the locations of sparse corruptions. In particular, we assume that the observation indicators satisfy
\begin{equation}\label{eq01}
  Z_{ij} = \mathbf{1}\{(i,j)\ \text{is observed}\} \sim \mathrm{Bern}(p_{ij}),~~\text{independently for all}~(i, j)\in [d_1]\times[d_2],
\end{equation}
with nonzero sampling probabilities $p_{ij}$. The fully observed setting corresponds to $p_{ij} \equiv 1$; otherwise, the model allows for heterogeneous missingness. Given the observed entries $Y_{\Omega_{\mathrm{obs}}}=[Y_{ij}]_{(i,j)\in \Omega_{\mathrm{obs}}}$, our goal is to construct confidence intervals $\{\widehat{C}(i,j)\}$ for the unobserved low-rank entries $(i,j)\in \Omega_{\mathrm{obs}}^{\rm{c}}$ such that the average coverage satisfies
$$
\mathbb{E}[\mathrm{AvgCov}(\widehat{C}; X, \Omega_{\mathrm{obs}})] \ge 1-\alpha,
$$
where
\begin{equation}\label{eq02}
  \mathrm{AvgCov}(\widehat{C}; X, \Omega_{\mathrm{obs}}) = \frac{1}{|\Omega_{\mathrm{obs}}^{\rm{c}}|} \sum_{(i,j)\in\Omega_{\mathrm{obs}}^{\rm{c}}} \mathbb{1} \{X_{ij} \in \widehat{C}(i,j)\}
\end{equation}
denotes the average coverage rate over unobserved entries $\Omega_{\mathrm{obs}}^{\rm{c}}$. By convention, when $\Omega_{\mathrm{obs}}^{\rm{c}}=\varnothing$, we set 
$$
\mathrm{AvgCov}(\widehat{C};X,\Omega_{\mathrm{obs}}) \equiv 1.
$$ 
This formulation aligns naturally with the conformal inference framework, enabling finite-sample, distribution-free uncertainty quantification for RPCA under arbitrary noise distributions, rank misspecification, heterogeneous missingness and potentially adversarial sparse corruptions.

\subsection{Computationally efficient two-stage RPCA via split CP}

In this section, we present the two-stage CP-RPCA algorithm by integrating CP with RPCA to quantify uncertainty in low-rank matrix recovery. By conformalizing the output of an arbitrary RPCA estimator, the proposed framework yields entrywise confidence intervals for the low-rank component with finite-sample coverage guarantees, without imposing distributional assumptions on the noise or relying on a specific recovery algorithm. Due to noise, sparse corruptions and potential model misspecification, RPCA estimators generally incur non-negligible estimation errors. To account for this uncertainty, we adopt a split CP strategy to construct confidence intervals of the form
\be
\widehat{C}(i,j) = \widehat{X}_{ij} \pm \widehat{q}\,\widehat{\sigma}_{ij},~~~(i,j)\in \Omega_{\mathrm{obs}}^{\rm{c}},
\ee
where $\widehat{X}_{ij}$ denotes the estimated low-rank entry, $\widehat{\sigma}_{ij}$ is an entrywise uncertainty scale and $\widehat{q}$ is a data-driven calibration constant.

\begin{figure}
	\centering
		\includegraphics[scale=0.9]{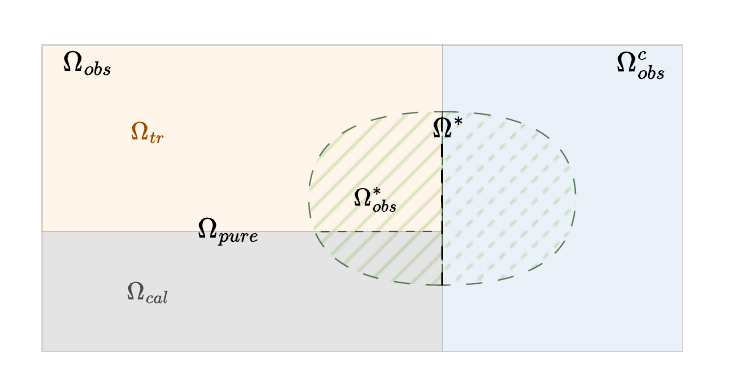}
	\caption{Relationships among index sets in the two-stage CP-RPCA framework}
	\label{FIG:1}
\end{figure}

Distinct from \cite{gui2023conformalized},  our proposed CP-RPCA employs a two-stage procedure to generate confidence intervals for the low-rank component. In the first stage, the observation indicator matrix $Z=[Z_{ij}]$ is generated according to the observation probabilities ${p_{ij}}$, which partitions the index set into the observed set $\Omega_{\mathrm{obs}}$ and the target set $\Omega_{\mathrm{obs}}^c$. Following the split conformal paradigm, the observed set is further divided into a training set $\Omega_{\mathrm{tr}}$ and a calibration set $\Omega_{\mathrm{cal}}$ with $\Omega_{\mathrm{obs}} = \Omega_{\mathrm{tr}} \cup \Omega_{\mathrm{cal}}$.\footnote[1]{When no confusion arises, we omit the subscript ``obs'' for notational simplicity.} Recall that $\Omega^{\ast}$ denotes the support set of the corruption matrix $S^{\ast}$. We further denote by $\Omega_{\mathrm{obs}}^{\ast}\subseteq \Omega_{\mathrm{obs}}$ the subset of observed indices corresponding to corrupted entries. The set of observed indices that are free of corruption is denoted by 
$$
\Omega_{\mathrm{pure}}: = \Omega_{\mathrm{obs}}\setminus \Omega_{\mathrm{obs}}^{\ast}.
$$ 
For clarity, the relationships among the different index sets are illustrated in Figure~\ref{FIG:1}. 

In the first stage, the training data are used to estimate the low-rank matrix $\widehat{X}$, the entrywise uncertainty scale $\widehat{\sigma}_{ij}$, the observation probability matrix $\widehat{P}$ and the set $\widehat{\Omega}_{\mathrm{obs}}^{\ast}$ of indices corresponding to potentially corrupted observations. Using this information, the calibration set is refined by removing entries identified as potentially corrupted, yielding a trimmed calibration set
$$
\Omega_{\mathrm{cal}}^{\prime}=\Omega_{\mathrm{cal}}\setminus \widehat{\Omega}_{\mathrm{obs}}^{\ast},
$$
where $\Omega_{\mathrm{cal}}^{\prime}$ ideally contains entries satisfying $Y_{ij}=X_{ij}$. 

In the second stage, standardized residual scores are computed over $\Omega_{\mathrm{cal}}^{\prime}$ and the calibration threshold $\widehat{q}$ is obtained via empirical quantiles to ensure that the resulting confidence intervals achieve the target coverage level $1-\alpha$. By explicitly excluding suspected outliers from calibration, this two-stage design improves robustness and yields more reliable uncertainty quantification than single-stage conformalization schemes.

\begin{algorithm}[!t]
\caption{Conformal Prediction Robust PCA (CP-RPCA)}
\label{alg:CPRPCA}
\begin{algorithmic}[1]
\Require target coverage level $1-\alpha$; data splitting proportion $q \in (0,1)$; observed entries $Y_{\Omega_{\mathrm{obs}}}$.

\State Split the data in the observation set: draw $W_{ij}\overset{\text{i.i.d.}}{\sim}\mathrm{Bern}(q)$ and define training and calibration sets
$$
\Omega_{\mathrm{tr}} = \{(i,j)\in \Omega_{\mathrm{obs}}: W_{ij}=1\}, 
~~\text{and}~~
\Omega_{\mathrm{cal}} = \{(i,j)\in \Omega_{\mathrm{obs}}: W_{ij}=0\}.
$$

\State Using the training data $Y_{\Omega_{\mathrm{tr}}}$:
\begin{itemize}
    \item Use an RPCA algorithm to estimate the low-rank component $\widehat{X}$, along with the index set $\widehat{\Omega}_{\mathrm{obs}}^{\ast}$ corresponding to potentially corrupted observations. The calibration set is then trimmed by removing these indices, yielding the refined calibration set $\Omega_{\mathrm{cal}}^{\prime}=\Omega_{\mathrm{cal}}\setminus \widehat{\Omega}_{\mathrm{obs}}^{\ast}$. Accordingly, the estimated set of observed indices that are free of corruption is defined as $\widehat{\Omega}_{\mathrm{pure}} = \Omega_{\mathrm{obs}}\setminus \widehat{\Omega}_{\mathrm{obs}}^{\ast}$.
    
    \item Use the Bootstrap method to compute $\widehat{\sigma}^2_{\mathrm{model}}$.  
    Calculate the noise variance $\widehat{\sigma}^2_{\mathrm{noise}}$ on the uncontaminated data in the training set: $\widehat{\sigma}_{\mathrm{noise}}^{2}=\lVert Y_{\widehat{\Omega}_{\mathrm{pure}}}-\widehat{X}_{\widehat{\Omega}_{\mathrm{pure}}} \rVert _{F}^{2}/ |\widehat{\Omega}_{\mathrm{pure}}|$,
    and thereby compute the relative uncertainty $\widehat{\sigma}_{ij}^{2}$ of $\widehat{X}_{ij}$ by $\widehat{\sigma}_{ij}^{2}=\widehat{\sigma}_{\mathrm{model}}^{2}+\widehat{\sigma}_{\mathrm{noise}}^{2}$;

    \item Compute the estimated observation probability matrix $\widehat{P}$, where $\widehat{p}_{ij}$ denotes the probability that entry $(i,j)$ is observed; the corresponding estimation procedure is described in Section~\ref{simulation}.
\end{itemize}

\State Compute the standardized residual scores for the uncontaminated entries in the calibration set: $\widehat{R}_{ij} = |Y_{ij}-\widehat{X}_{ij}|/\widehat{\sigma}_{ij}$ for $(i,j)\in \Omega_{\mathrm{cal}}^{\prime}$.

\State Compute estimated odds ratios for the calibration set and test set: $\widehat{h}_{ij} = (1-\widehat{p}_{ij})/\widehat{p}_{ij}$, and then calculate the weights for the calibration set and test point,
\begin{equation}\label{alg1_eq01}
  \widehat{\omega}_{ij}=\frac{\widehat{h}_{ij}}{\sum_{(i^{\prime},j^{\prime})\in\Omega_{\mathrm{cal}}^{\prime}} \widehat{h}_{i^{\prime}j^{\prime}}
    + \max_{(i^{\prime},j^{\prime})\in\Omega_{\mathrm{obs}}^c} \widehat{h}_{i^{\prime}j^{\prime}}},~~(i,j)\in\Omega_{\mathrm{cal}}^{\prime},
    ~~~
  \widehat{\omega}_{\mathrm{test}}=\frac{\max_{(i,j)\in\Omega_{\mathrm{obs}}^c} \widehat{h}_{ij}}{\sum_{(i^{\prime},j^{\prime})\in\Omega_{\mathrm{cal}}^{\prime}} \widehat{h}_{i^{\prime}j^{\prime}} + \max_{(i^{\prime},j^{\prime})\in\Omega_{\mathrm{obs}}^c} \widehat{h}_{i^{\prime}j^{\prime}}}.
\end{equation}

\State Compute threshold $\widehat{q} = \mathrm{Quantile}_{1-\alpha}\big(
\sum_{(i,j)\in \Omega_{\mathrm{cal}}^{\prime}}\widehat{\omega}_{ij}\cdot\delta_{\widehat{R}_{ij}} + \widehat{\omega}_{\mathrm{test}}\cdot\delta_{+\infty}\big)$.

\Ensure Confidence intervals $\widehat{C}(i,j)=\widehat{X}_{ij} \pm \widehat{q} \cdot \widehat{\sigma}_{ij}$ for each unobserved entry.
\end{algorithmic}
\end{algorithm}

{\remark
The CP-RPCA framework departs from classical RPCA by removing distributional assumptions on the noise $E$. Following the perspective of \cite{sun2025conformalized}, our approach preserves the low-rank plus sparse decomposition: the low-rank component provides a compact representation for dimension reduction and compression, while the sparse component explicitly captures and localizes outliers or sparse corruptions, enabling robust recovery and anomaly detection. Moreover, motivated by \cite{gui2023conformalized} and by the fact that the low-rank component often represents a stable background structure in practical applications (see Section~\ref{sec5.2}), we treat it as a deterministic matrix when performing uncertainty quantification. In particular, we do not require independence, Gaussianity, or moment conditions. Instead, all randomness arises from the sampling of observed indices and the (possibly adversarial) locations of sparse corruptions. We assume only that the observed matrix $Y$ can be decomposed as the sum of a sparse component $S^{\ast}$ and an approximately low-rank component $X$.
}

\subsection{Exact two-stage RPCA via full CP}

For completeness, we present a fully conformal version of CP-RPCA (see Algorithm~\ref{alg:full_cprpca}). Compared with the split conformal approach, this variant avoids sample splitting and therefore provides slightly tighter finite-sample guarantees, at the expense of increased computational cost. Given the observed matrix $Y_{\Omega_{\mathrm{obs}}}$ and a test point $(i_{\ast},j_{\ast}) \in \Omega_{\mathrm{obs}}^c$, consider a candidate value $y \in \mathcal{Y}$ for the unobserved entry $Y_{i_{\ast}j_{\ast}}$, where $\mathcal{Y}$ denotes the admissible value space of the missing entries. Define the augmented matrix $Y^{(y)}$ entrywise by
$$
Y_{ij}^{(y)}=
\begin{cases}
  Y_{ij}, & (i,j)\in\Omega_{\mathrm{obs}},\\
  y, & (i,j)=(i_\ast,j_\ast),\\
  \emptyset, & \text{otherwise},
\end{cases}
$$
where $Y_{ij}^{(y)}=\emptyset$ indicates that the corresponding entry is unobserved. 

\begin{algorithm}[t]
\caption{Full CP-RPCA: A Fully Conformal Inference Procedure for Robust PCA}
\label{alg:full_cprpca}
\begin{algorithmic}[1]
\Require target coverage level $1-\alpha$, observed entries $Y_{\Omega_{\mathrm{obs}}}$.

\State Compute the observation probability matrix $\widehat{P}$ using the observed data $Y_{\Omega_{\mathrm{obs}}}$, where $\widehat{p}_{ij}$ estimates the probability that entry $(i,j)$ is observed; the specific estimation procedure for $p_{ij}$ is described in Section~\ref{simulation}.

\For{$(i_\ast,j_\ast)\in\Omega_{\mathrm{obs}}^c$}
    \For{$y\in\mathcal{Y}$}

        \State Add the assumed data to the observed set $Y_{\Omega_{\mathrm{obs}}}$ to obtain the augmented dataset $Y^{(y)}$. Using the augmented dataset $Y^{(y)}$:

        \begin{itemize}
            \item Apply Algorithm~\ref{alg:fast_rpca} to compute the estimated matrix $\widehat{X}^{(y)}$ and estimate the set of uncontaminated element indices $\widehat{\Omega}_{\mathrm{pure}}^{(y)}$.
            
            \item Use Bootstrap to compute $\widehat{\sigma}^{2}_{\mathrm{model}}$ and calculate the noise variance $\widehat{\sigma}^{2}_{\mathrm{noise}}$ on the uncontaminated training data: 
                $
                \widehat{\sigma}^{2}_{\mathrm{noise}} = {\|Y_{\widehat{\Omega}_{\mathrm{pure}}^{(y)}} - \widehat{X}_{\widehat{\Omega}_{\mathrm{pure}}^{(y)}}\|_{F}^{2}}/{|\widehat{\Omega}_{\mathrm{pure}}^{(y)}|}
                $, 
                and compute the relative uncertainty: 
                $
                \widehat{\sigma}_{ij}^{2}=\widehat{\sigma}^{2}_{\mathrm{model}}+\widehat{\sigma}^{2}_{\mathrm{noise}}
                $.
        \end{itemize}

        \State Calculate the standardized residual scores for the uncontaminated entries in the observation set and the test points:
        $
        \widehat{R}^{(y)}_{ij} = {|Y^{(y)}_{ij}-\widehat{X}^{(y)}_{ij}|}/{\widehat{\sigma}^{2}_{ij}},~~(i,j)\in \widehat{\Omega}_{\mathrm{pure}}^{(y)}\cup\{(i_\ast,j_\ast)\}
        $.
        
        \State Compute the estimated odds ratio for the observation set and test points: $\widehat{h}_{ij}={(1-\widehat{p}_{ij})}/{\widehat{p}_{ij}}$ for $(i,j)\in\widehat{\Omega}_{\mathrm{pure}}^{(y)}\cup \{(i_\ast,j_\ast)\}$, and subsequently compute the weights $\widehat{\omega}_{ij}$ for calibration and test data using Eq.~\eqref{alg1_eq01}.

        \State Calculate the weighted quantile:
        $
        \widehat{q}^{(y)}(i_\ast,j_\ast)
        = \mathrm{Quantile}_{1-\alpha}\big(\sum_{(i,j)\in\widehat{\Omega}_{\mathrm{pure}}^{(y)}} \widehat{\omega}_{ij}\delta_{\widehat{R}^{(y)}_{ij}} + \widehat{\omega}_{i_\ast j_\ast}\delta_{\widehat{R}^{(m)}_{i_\ast j_\ast}}\big)
        $.
        
    \EndFor
\EndFor

\Ensure $\big\{\widehat{C}(i_{\ast},j_{\ast}) = \{y\in \mathcal{Y}: \widehat{R}_{i_{\ast}j_{\ast}}^{(y)} \le \widehat{q}^{(y)}(i_{\ast},j_{\ast})\}: (i_{\ast},j_{\ast}) \in \Omega_{\mathrm{obs}}^c \big\}$.

\end{algorithmic}
\end{algorithm}

\section{Theoretical results}

\subsection{Weighted exchangeability}

In Algorithm~\ref{alg:CPRPCA}, we adopt a split conformal inference scheme by partitioning the observed index set $\Omega_{\mathrm{obs}}$ into a training set $\Omega_{\mathrm{tr}}$ and a calibration set $\Omega_{\mathrm{cal}}$, which induces dependence among the resulting entries. In addition, heterogeneous observation probabilities $p_{ij}$ lead to a distributional shift relative to the i.i.d. setting assumed in standard CP.  Let 
$$
\Omega_{\mathrm{cal}}^{\prime} \cup \{(i_\ast,j_\ast)\} {:=} \{(i_1,j_1),\ldots,(i_{n_{\mathrm{cal}}+1},j_{n_{\mathrm{cal}}+1})\},
$$
where $(i_\ast,j_\ast)$ denotes a test point and $n_{\mathrm{cal}}=|\Omega_{\mathrm{cal}}^{\prime}|$. When $p_{ij}$ is non-constant, the test point cannot appear with equal probability at each of the $n_{\mathrm{cal}}+1$ positions and hence classical exchangeability fails. 
To ensure the validity of conformal inference, we show that the weighted exchangeability holds as follows.
\begin{lemma}\label{lem01}
If $(i_\ast,j_\ast)\mid\Omega_{\mathrm{obs}}\sim\mathrm{Unif}(\Omega_{\mathrm{obs}}^c)$, then
\begin{equation*}
\mathbb{P}\big\{(i_\ast,j_\ast) = (i_k,j_k) \mid \Omega_{\mathrm{cal}}^{\prime}\cup\{(i_\ast,j_\ast)\} = \{(i_1,j_1),\ldots,(i_{n_{\mathrm{cal}}+1},j_{n_{\mathrm{cal}}+1})\}, \Omega_{\mathrm{tr}}\big\} = \omega_{i_k j_k},
\end{equation*}
where we define the weight $\omega_{i_k j_k}=h_{i_kj_k}/\sum_{k^{\prime}=1}^{n_{\mathrm{cal}}+1}{h_{i_{k^{\prime}}j_{k^{\prime}}}}$ for odds ratios given by $h_{ij}=(1-p_{ij})/p_{ij}$.
\end{lemma}

Lemma 1 indicates that we can correct the distribution shift induced by non-uniform observation probabilities. Following \cite{tibshirani2019conformal}, this yields a weighted exchangeability framework under which the test-point score behaves as if it were drawn from a reweighted empirical distribution, i.e.,
$$\widehat{R}_{i_k j_k}\mid \Omega_{\mathrm{cal}}^{\prime}\cup \Omega_{\mathrm{tr}} \cup \{(i_\ast,j_\ast)\} \sim \sum_{k=1}^{n_{\mathrm{cal}}+1}{\omega}_{i_k j_k}\delta_{R_{i_k j_k}}.
$$
Consequently, the following result holds.
\begin{corollary}
Suppose the observation probability matrix $P$ is known and the weights ${\omega}_{i_k j_k}$ are correctly specified. Then for any symmetric score function $R$, and for any test point $(i_\ast,j_\ast)\mid\Omega_{\mathrm{obs}}\sim\mathrm{Unif}(\Omega_{\mathrm{obs}}^c)$, then
$$
\mathbb{P}\{X_{i_{\ast}j_{\ast}}\in \widehat{C}(i_{\ast}, j_{\ast})\}
=\mathbb{P}\{X_{i_{\ast}j_{\ast}}\in \widehat{X}_{i_{\ast}j_{\ast}}\pm q_{i_{\ast}j_{\ast}}^{\ast}\cdot \widehat{\sigma}_{i_{\ast}j_{\ast}}\}
=\mathbb{P}\{ R_{i_{\ast}j_{\ast}}\le q_{i_{\ast}j_{\ast}}^{\ast}\}
\ge 1-\alpha,
$$
where $q_{i_{\ast}j_{\ast}}^{\ast}=\mathrm{Quantile}_{1-\alpha}\big(\sum_{k=1}^{n_{\mathrm{cal}}+1}\omega_{i_k j_k}\delta_{R_{i_k j_k}}\big)$ is the adjusted threshold.
\end{corollary}
\noindent The threshold $q_{i_{\ast}j_{\ast}}^{\ast}$ generally depends on the specific test point $(i_{\ast},j_{\ast})$, which would require recalibration for each unobserved entry. To reduce the resulting computational burden, we adopt the one-shot weighting strategy of \citet{gui2023conformalized}, which constructs a unified upper bound applicable to all test points:
$$
q_{i_{\ast}j_{\ast}}^{\ast} \leq q^{\ast} = \mathrm{Quantile}_{1-\alpha}\Big(\sum_{(i,j)\in \Omega_{\mathrm{cal}}^{\prime}}\omega_{ij}^{\ast}\delta_{S_{ij}}+\omega_{\mathrm{test}}^{\ast}\,\delta_{+\infty}\Big).
$$
Here, $q^{\ast}$ serves as a common upper bound for $\{ q_{i_{\ast}j_{\ast}}^{\ast}: (i_{\ast},j_{\ast}) \in \Omega_{\mathrm{obs}}^c\}$, with weights defined by
$$
\omega_{ij}^{\ast} = \frac{h_{ij}}{\sum_{(i^{\prime},j^{\prime})\in\Omega_{\mathrm{cal}}^{\prime}} h_{i^{\prime}j^{\prime}} + \max_{(i^{\prime},j^{\prime})\in\Omega_{\mathrm{obs}}^c} h_{i^{\prime}j^{\prime}}},~~~
\omega_{\mathrm{test}}^{\ast} = \frac{\max_{(i,j)\in\Omega_{\mathrm{obs}}^c} h_{ij}}{\sum_{(i^{\prime},j^{\prime})\in\Omega_{\mathrm{cal}}^{\prime}} h_{i^{\prime}j^{\prime}} + \max_{(i^{\prime},j^{\prime})\in\Omega_{\mathrm{obs}}^c} h_{i^{\prime}j^{\prime}}}.
$$
The term $\omega_{test}^{\ast}\delta _{+\infty} $ acts as a finite-sample correction to ensure valid coverage. As a result, the calibrated threshold $q^{\ast}$ no longer depends on the specific test point. In practice, the observation probabilities $p_{ij}$ are unknown and must be estimated. Replacing $p_{ij}$ with their estimates $\widehat{p}_{ij}$, we compute the corresponding weights as specified in equation (\ref{alg1_eq01}) of Algorithm~\ref{alg:CPRPCA}. The conclusion of weighted exchangeability under the full conformal CP-RPCA framework is omitted since it similar to the split conformal case. 

\subsection{Theoretical guarantee}

In practice, both the observation probabilities $p_{ij}$ and the index set $\Omega_{\mathrm{obs}}^{\ast}$ corresponding to corrupted observations are unknown. We therefore perform calibration using the estimated observation probabilities $\widehat{p}_{ij}$ and the estimated trimmed calibration set $\Omega_{\mathrm{cal}}^{\prime}=\Omega_{\mathrm{cal}}\setminus \widehat{\Omega}_{\mathrm{obs}}^{\ast}$, which leads to two sources of error analogous to \cite{gui2023conformalized}. 
The first source of error arises from replacing the true observation probabilities $p_{ij}$ with their estimates $\widehat{p}_{ij}$, which induces a discrepancy between the ideal calibration weights and their empirical counterparts. To capture this effect, we define the total variation distance 
\begin{equation}
\label{eq04}
\Delta =\frac{1}{2}\sum_{(i,j) \in \Omega_{\text{cal}}^{\prime}\cup\{(i_{\ast},j_{\ast})\}}{\Bigg|\frac{\widehat{h}_{ij}}{\sum_{(i^{\prime},j^{\prime}) \in \Omega_{\text{cal}}^{^{\prime}}\cup\{( i_{\ast},j_{\ast})\}}{\widehat{h}_{i^{\prime}j^{\prime}}}} - \frac{h_{ij}}{\sum_{(i^{\prime},j^{\prime})\in\Omega_{\mathrm{cal}}^{^{\prime}}\cup\{(i_{\ast},j_{\ast})\}}{h_{i^{\prime}j^{\prime}}}}\Bigg|},
\end{equation}
where $(i_{\ast},j_{\ast})$ is the test point. As in \cite{gui2023conformalized}, $\Delta$ quantifies the extent to which weight estimation error perturbs the conformal calibration step and directly contributes to coverage distortion. The second source of error is due to imperfect identification of uncontaminated entries. Because $\widehat{\Omega}_{\mathrm{obs}}^{\ast}$ is estimated, the trimmed calibration set $\Omega_{\mathrm{cal}}^{\prime}$ may still contain entries corrupted by sparse outliers, which  results in a distributional mismatch between the empirical score distribution used for calibration and its oracle counterpart. Specifically, let
$$
\big\{\widehat{R}_{i_1 j_1}, \cdots ,\widehat{R}_{i_l j_l}, \widehat{R}_{i_{l+1} j_{l+1}}, \cdots ,\widehat{R}_{i_N j_N}\big\}~~\text{and}~~\big\{R_{i_1 j_1}, \cdots, R_{i_l j_l}, R_{i_{l+1} j_{l+1}}, \cdots, R_{i_N j_N}\big\}
$$ 
denote the empirical and oracle score vectors on $\Omega_{\mathrm{cal}}^{\prime}$, respectively, where
$
\widehat{R}_{ij}={|Y_{ij}-\widehat{X}_{ij}|}/{\widehat{\sigma}_{X_{ij}}}$, $R_{ij}={|X_{ij}-\widehat{X}_{ij}|}/{\widehat{\sigma}_{X_{ij}}},
$
$l=|\Omega_{\mathrm{cal}}^{\prime} \cap \Omega_{\mathrm{obs}}^{\ast}|$, $N=|\Omega_{\mathrm{cal}}^{\prime}|$, $\widehat{R}_{i_k j_k} \ne R_{i_k j_k}$ for $1\le k\le l$ and $\widehat{R}_{i_k j_k}=R_{i_k j_k}$ for $l+1\le k\le N$. We summarize the resulting discrepancy by
\begin{equation}\label{eq05}
  \xi = \sum_{\{{(i,j)\in\Omega_{\mathrm{cal}}^{\prime}: \widehat{R}_{ij}\neq R_{ij}}\}} \widehat{\omega}_{ij},
\end{equation}
where $\xi$ quantifies the error caused by the distribution shift due to the presence of contaminated entries in the data used for calibration. In addition, we make the following assumptions:

\begin{assumption}[Vanishing maximum calibration weight]
Let $\omega_{\max}=\max\{\omega_{ij}: (i,j) \in \Omega^{\prime}_{\mathrm{cal}} \cup \Omega_{\mathrm{obs}}^c\}$, we assume $\omega_{\max}=o(1)$ as $N\rightarrow \infty$. 
\end{assumption}

\begin{assumption}[Continuity of the residual score distribution]
The distribution of the residual scores is continuous, so that ties occur with probability zero.
\end{assumption}

\begin{assumption}[Negligibility of one-shot weighting error]
The additional error introduced by the one-shot weighting scheme in the coverage guarantee is asymptotically negligible as $N\rightarrow \infty$.
\end{assumption}

{\remark The above assumptions are imposed to establish the theoretical results in the subsequent section. Assumption~1 is a mild regularity condition required to control the upper bound in the coverage analysis. It ensures that no single calibration or target index receives a dominant weight. In particular, under uniform sampling, one has $\omega_{\max}=(N+1)^{-1}$, which trivially satisfies this condition. Assumption~2 is standard in the CP literature (see, e.g., \citealp{lei2018distribution}) and guarantees the uniqueness of empirical quantiles. This assumption can be removed by employing randomized tie-breaking, at the cost of additional notational complexity. Assumption~3 concerns the use of one-shot weighting. When one-shot weighting is employed, the calibration weight is given by $\widehat{\omega}_{ij}={\widehat{h}_{ij}}/({\sum_{(i^{\prime},j^{\prime}) \in \Omega_{\mathrm{cal}}^{\prime}}\widehat{h}_{i^{\prime}j^{\prime}}+\max_{(i^{\prime},j^{\prime}) \in \Omega_{\mathrm{obs}}^c}\widehat{h}_{i^{\prime}j^{\prime}}})$, whereas the ideal weight without one-shot weighting takes the form ${\omega}_{ij}^{\prime}={\widehat{h}_{ij}}/({\sum_{(i^{\prime},j^{\prime}) \in \Omega_{\mathrm{cal}}^{\prime}\cup \{(i_{\ast},j_{\ast})\}}{\widehat{h}_{i^{\prime}j^{\prime}}}})$. When the trimmed calibration set $\Omega_{\mathrm{cal}}^{\prime}$ is sufficiently large, the additional term $\max_{(i^{\prime},j^{\prime}) \in \Omega_{\mathrm{obs}}^c}\widehat{h}_{i^{\prime},j^{\prime}}$ is negligible relative to $\sum_{(i^{\prime},j^{\prime}) \in \Omega_{\mathrm{cal}}^{\prime}}{\widehat{h}_{i^{\prime},j^{\prime}}}$. As a result, the discrepancy between the two denominators vanishes asymptotically, implying that the coverage error induced by one-shot weighting can be safely ignored.}

Now we have the following coverage guarantee:
\begin{theorem}\label{th01}
Let $\widehat{X}$, $\widehat{\sigma}_{ij}$ and  $\widehat{p}_{ij}$ denote consistent estimators of $X$, $\sigma_{ij}$ and  $p_{ij}$, respectively,  depending on the data solely through the training sample $Y_{\Omega_{\mathrm{tr}}}$. Under Assumptions~1--3, the CP-RPCA confidence intervals satisfy
\begin{equation*}
    \mathbb{E}[\operatorname{AvgCov}(\widehat{C}; X, \Omega_{\mathrm{obs}})] \ge 1-\alpha-\mathbb{E}[\Delta] -\mathbb{E}[\xi].
\end{equation*}
Further assume that the score $\widehat{R}_{ij}$ for $(i,j) \in \Omega_{\mathrm{cal}}^{\prime}$ has a continuous distribution, then there is a coverage upper bound:
\begin{equation*}
\begin{aligned}
\mathbb{E}[\operatorname{AvgCov}(\widehat{C}; X,\Omega_{\mathrm{obs}})]
   &\le 1-\alpha +2\omega_{\max}+\mathbb{E}[\Delta]+\mathbb{E}[\xi]\\
   &= 1-\alpha + o(1) + \mathbb{E}[\Delta]+\mathbb{E}[\xi].
\end{aligned}
\end{equation*}
\end{theorem}
\noindent These bounds ensure that the proposed intervals are neither under-covering nor excessively conservative. Based on Theorem~\ref{th01}, we can get the following corollaries:
\begin{corollary}
When the observation probability $p_{ij}$ of each entry is known, we have the following coverage guarantee:
$$
\mathbb{E}[\operatorname{AvgCov}(\widehat{C}; X, \Omega_{\mathrm{obs}})] \le 1-\alpha +2\omega_{\max}+\mathbb{E}[\xi].
$$
Furthermore, under uniform sampling with $p_{ij}\equiv~\text{constant}~p_0$ and $\omega_{ij}=1/(N+1)$, the coverage guarantee holds:
$$
\mathbb{E}[\operatorname{AvgCov}(\widehat{C}; X, \Omega_{\mathrm{obs}})] \le 1-\alpha +\frac{2}{N+1}+\mathbb{E}[\xi].
$$
\end{corollary}

\begin{corollary}
When the positions of the outliers are known, the RPCA problem reduces to a matrix completion problem and $\xi=0$. In this case, the coverage guarantee is given by:
$$
1-\alpha -\mathbb{E}[\Delta] \leq \mathbb{E}[\operatorname{AvgCov}(\widehat{C}; X, \Omega_{\mathrm{obs}})] \le 1-\alpha +2\omega _{\max}+\mathbb{E}[\Delta].
$$
\end{corollary}

Regarding Theorem~\ref{th01}, we provide the following additional clarifications.
\begin{remark}
The coverage guarantee established above corresponds to marginal (or average) coverage and should be distinguished from conditional coverage. Conditional coverage requires the prescribed coverage level to hold at every individual test point, which is generally unattainable with finite-length prediction intervals without imposing substantially stronger assumptions; see \cite{lei2014classification} for a formal discussion.
\end{remark}

\begin{remark}
The error term $\xi$ is defined in a conservative manner. In Theorem~\ref{th01}, it implicitly allows for an extreme scenario, where the contaminated scores $\widehat{R}_{ij}$ on $\Omega_{\mathrm{cal}}^{\prime} \cap \Omega_{\mathrm{obs}}^{\ast}$ ($\Omega_{\mathrm{cal}}^{\prime}=\Omega_{\mathrm{cal}}\setminus \widehat{\Omega}_{\mathrm{obs}}^{\ast}$) undergo a complete reversal in their relative ordering compared to the oracle scores $R_{ij}$ (e.g., all oracle scores fall below the threshold $q$, while all contaminated scores exceed it). In practice, however, the discrepancy between $\widehat{R}_{ij}$ and $R_{ij}$ is typically small, leading only to mild perturbations in the score ranking and a correspondingly limited impact on the empirical coverage.
\end{remark}

\begin{theorem}\label{th02}
Let $\widehat{X}^{(y)}$, $\widehat{\sigma}_{ij}$ and $\widehat{p}_{ij}$ be consistent estimators of $X$, $\sigma_{ij}$ and  $p_{ij}$, respectively, constructed using only the augmented matrix $Y^{(y)}$. Then the confidence intervals $\widehat{C}$ produced by Algorithm~\ref{alg:full_cprpca} satisfy the following lower bound
$$
\mathbb{E}[\operatorname{AvgCov}(\widehat{C}; X, \Omega_{\mathrm{obs}})] \ge 1-\alpha -\mathbb{E}[\Delta] -\mathbb{E}[\xi]. 
$$
If, in addition, the score variables $\widehat{R}_{ij}^{(y)}$, for $(i,j) \in \widehat{\Omega}_\mathrm{pure} \cup \{(i_{\ast},j_{\ast})\}$, have a continuous distribution, then the following upper bound holds:
$$
\mathbb{E}[\operatorname{AvgCov}(\widehat{C}; X, \Omega_{\mathrm{obs}})] \le 1-\alpha +\omega _{\max}+\mathbb{E}[\Delta] +\mathbb{E}[\xi],  
$$
where $\omega_{\max}$ denotes the maximum weight over $\big\{\omega_{ij}, (i,j) \in \widehat{\Omega}_\mathrm{pure} \cup \{(i_{\ast},j_{\ast})\}\big\}$.
\end{theorem}

\begin{corollary}
Under uniform sampling with $p_{ij}\equiv~\text{constant}~p_0$, the coverage bound simplifies to
$$
\mathbb{E}[\operatorname{AvgCov}(\widehat{C}; X, \Omega_{\mathrm{obs}})] \le 1-\alpha +\frac{1}{N+1}+\mathbb{E}[\xi], 
$$
where $N := |\Omega_{\mathrm{pure}}| = |\Omega_{\mathrm{obs}}\setminus \Omega_{\mathrm{obs}}^{\ast}|$ is the cardinality of the set of observed indices corresponding to uncontaminated entries.
\end{corollary}

\newpage
\section{Simulation studies}\label{simulation}

In this section, we conduct comprehensive simulation studies to evaluate the finite-sample performance of the proposed CP-RPCA method focusing  on the average coverage rate $\operatorname{AvgCov}(\widehat{C})$ defined in \eqref{eq02} together with the average interval length
$$
\operatorname{AvgLength}(\widehat{C}) = \frac{1}{|\Omega_{\mathrm{obs}}^c|}\sum_{(i,j)\in\Omega_{\mathrm{obs}}^c}\text{length}\bigl(\widehat{C}(i,j)\bigr).
$$
All results reported in this section are fully reproducible using the code provided in the supplementary materials.

\paragraph{Simulation Setup}
The data matrix is generated from
$$
Y=X^{\ast}+S^{\ast}+E,
$$
where $X^{\ast}$ is a rank-$r^{\ast}$ matrix with orthonormal factors $U^{\ast}\in \mathbb{R}^{d_1\times r^{\ast}}$ and $V^{\ast}\in \mathbb{R}^{d_2\times r^{\ast}}$, whose entries are drawn i.i.d. from a specified distribution $\mathcal{P}$ (specified below). The sparse corruption matrix $S^{\ast}$ has sparsity level $\beta$ and  $E$ denotes additive noise with distribution varying across experiments. The observation set $\Omega_{\mathrm{obs}}$ is generated according to the observation model in Eq.~\eqref{eq01}. We set the matrix dimensions $d_1=d_2=500$, the true rank  $r^{\ast}=8$, the training proportion $q=0.7$ and the nominal coverage level  $1-\alpha =0.90$. Model uncertainty $\widehat{\sigma}_{\mathrm{model}}^{2}$ is estimated via bootstrap with $50$ resamples. In each run, we deliberately vary the assumed rank $r$ and tune algorithmic parameters via cross-validation to evaluate robustness under rank misspecification. All simulation results are averaged over $50$ independent replications.

\subsection{Choice of RPCA algorithms}

Owing to the model-agnostic nature of CP-RPCA, any valid RPCA solvers can be used as the underlying estimation algorithm. Table~\ref{table1} compares the performance of CP-RPCA when paired with the Fast-RPCA algorithm of \cite{yi2016fast} and the augmented Lagrange multiplier (ALM) method of \cite{candes2011robust}, under uniform missingness and Gaussian noise. In this comparison, the maximum number of iterations and the stopping criteria are kept identical across algorithms. In Table~\ref{table1}, \textit{AvgTime} denotes the average runtime for a single execution on the observed data. 

As shown in Table~\ref{table1}, we have the following findings: (1) The coverage guarantee of CP-RPCA is insensitive to the choice of the underlying RPCA solver; the solver primarily affects the reconstruction accuracy and then influences the resulting interval lengths. (2) The ALM method does not require a prespecified rank $r$ and therefore yields a single set of performance metrics under each configuration, exhibiting relatively stable behavior. However, due to the repeated singular value decompositions required at each iteration, ALM is computationally expensive. Consequently, in all subsequent experiments, we adopt Fast-RPCA as the underlying algorithm. A detailed description of the Fast-RPCA procedure is provided in Appendix~\ref{appendix_01}.

\begin{table}[b]
\centering
\caption{Comparison of performance when using Fast-RPCA and ALM algorithms as the underlying methods}\label{table1}
\begin{tabular}{lccccc}
\toprule
& & \multicolumn{2}{c}{$\beta = 0.1$} & \multicolumn{2}{c}{$\beta = 0.3$} \\
\cline{3-6}
& & $r=4$ & $r=16$ & $r=4$ & $r=16$ \\
\hline
\multirow{3}{*}{\textbf{Fast-RPCA}} & AvgCov & 0.900 & 0.899 & 0.879 & 0.875 \\
                                    & AvgLength & 0.022 & 0.019 & 0.022 & 0.021 \\
                                    & AvgTime & 3.06 & 1.78 & 3.15 & 1.84 \\
\hline
\multirow{3}{*}{\textbf{RPCA-ALM}} & AvgCov & \multicolumn{2}{c}{0.899} & \multicolumn{2}{c}{0.879} \\
                                   & AvgLength & \multicolumn{2}{c}{0.022} & \multicolumn{2}{c}{0.023} \\
                                   & AvgTime & \multicolumn{2}{c}{26.10} & \multicolumn{2}{c}{25.90} \\
\bottomrule
\end{tabular}
\end{table}

\subsection{Robustness to missingness and noise distributions}

We begin by evaluating the finite-sample behavior of the proposed confidence intervals across different observation mechanisms and noise distributions with the sparsity level $\beta = 0.1$. The corresponding experimental configurations are described below.

\subsubsection{Homogeneous missingness:}

We consider two simulation settings under homogeneous missingness. In this case, the observation probabilities are constant, i.e., $p_{ij}\equiv p$, and are estimated by $\widehat{p}_{ij} = (d_1 d_2 q)^{-1}|\Omega_{\mathrm{tr}}|$ for all $(i,j)$.
\begin{itemize}
    \item \textbf{Setting 1:} Uniform observation + Gaussian noise, i.e.,  $p=0.5$, $\mathcal{P}_{u,v}=\mathcal{N}(0,1/r)$ and  $E_{ij} \sim \mathcal{N}\big(0, (0.1\lVert X^{\ast} \rVert_{\infty})^2\big)$.
    \item \textbf{Setting 2:} Uniform observation + heavy-tailed noise, i.e., $p=0.5$, $\mathcal{P}_{u,v}=\mathcal{N}(0,1/r)$ and  $E_{ij} \sim t(3)$.
\end{itemize}

\subsubsection{Heterogeneous missingness:}

Under heterogeneous missingness, the observed entries are no longer sampled uniformly. We consider two representative non-uniform observation mechanisms—logistic missingness and rank-one missingness.
\textbf{(1) Logistic Missing:} 
We generate $\{a_{il}: i \le d_1, l \le k^{\ast}\}$ i.i.d. from $\mathrm{Unif}(0,1)$ and $\{b_{lj}: l \le k^{\ast}, j \le d_2\}$ i.i.d. from $\mathrm{Unif}(-0.5,0.5)$. The observation probabilities are specified through the logistic model
$
\log({p_{ij}}/({1-p_{ij}})) = \sum_{l=1}^{k^{\ast}}{a_{il}b_{lj}}. 
$
When estimating the observation probability matrix, $\widehat{p}_{ij}$ is obtained by estimating $\widehat{a}_{il}$ and $\widehat{b}_{lj}$ via constrained maximum likelihood.
\textbf{(2) Rank-1 Missing:} 
We draw $a_i$ and $b_j$ independently from $\mathrm{Unif}(0.3,0.9)$ and define the observation probability as $p_{ij} = a_i b_j$.
The following Settings 3–6 combine heterogeneous missingness with Gaussian or heavy-tailed noise.
\begin{itemize}
    \item \textbf{Setting 3:} Logistic missing + Gaussian noise, where $k^{\ast}=5$ and  the noise distribution is the same as in Setting~1.
    \item \textbf{Setting 4:} Logistic missing + Heavy-tailed noise, where $k^{\ast}=5$ and  the noise distribution follows Setting~2.
    \item \textbf{Setting 5:} Rank-1 missing + Gaussian noise, where the noise distribution follows Setting~1.
    \item \textbf{Setting 6:} Rank-1 missing + Heavy-tailed noise, where the noise distribution follows Setting~2.
\end{itemize}

\begin{figure}
	\centering
		\includegraphics[scale=0.45]{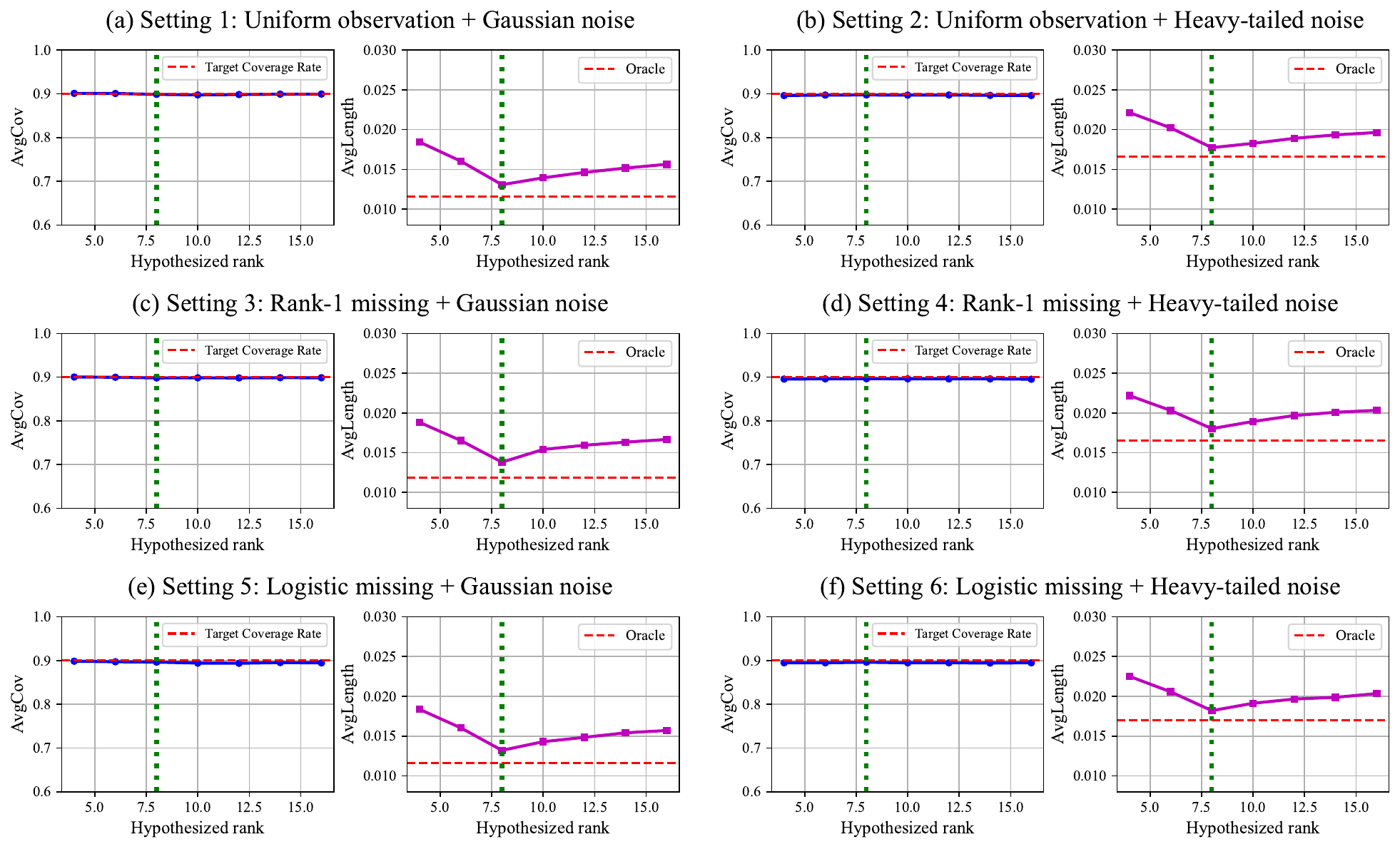}
	\caption{Comparison of coverage effects under different observation modes and noise distributions}
	\label{FIG:2}
\end{figure}

Figure~\ref{FIG:2} reports the empirical coverage and average interval length as the hypothesized rank $r$ varies from 4 to 16. Across all observation mechanisms and noise distributions, CP-RPCA achieves coverage close to the nominal level, demonstrating robustness to rank misspecification and data heterogeneity. When the rank is correctly specified ($r=r^{\ast}$), the resulting intervals are shortest and nearly approach the oracle length, defined as the difference between the $(1-\alpha/2)$-th and $(\alpha/2)$-th quantiles of the noise distribution. Under rank underestimation ($r<r^{\ast}$), unmodeled low-rank structure is absorbed into the sparse and noise components, inflating estimation uncertainty and leading to wider intervals. Conversely, rank overestimation ($r>r^{\ast}$) increases the estimated model error variance, again resulting in longer intervals to maintain valid coverage. Overall, these results confirm that CP-RPCA provides distribution-free, entrywise confidence intervals with stable coverage and adaptive length across a wide range of noise distributions, missingness mechanisms and  model misspecifications.

\subsection{Heterogeneous and adversarial noise}

We further examine the coverage performance of CP-RPCA under heterogeneous and adversarial noise by considering the following two settings:

\begin{itemize}
    \item \textbf{Setting 7:} Logistic missing + Adversarial heterogeneous noise, where $E_{ij} \sim \mathcal{N}\big(0, (0.1\lVert X^{\ast} \rVert_{\infty}\sigma_{ij})^2\big)$, $\sigma_{ij}=(2p_{ij})^{-1}$ and  high-variance noise is deliberately concentrated on entries that are least likely to be observed during training.
    \item \textbf{Setting 8:} Logistic missing + Random heterogeneous noise, where $E_{ij} \sim \mathcal{N}\big(0, (0.1\lVert X^{\ast} \rVert_{\infty}\sigma_{ij})^2\big)$ and  $\sigma_{ij}$ is randomly drawn from the empirical distribution of ${(2p_{ij})^{-1}}_{(i,j)\in[d_1]\times[d_2]}$.
\end{itemize}

\begin{figure}
	\centering
		\includegraphics[scale=0.45]{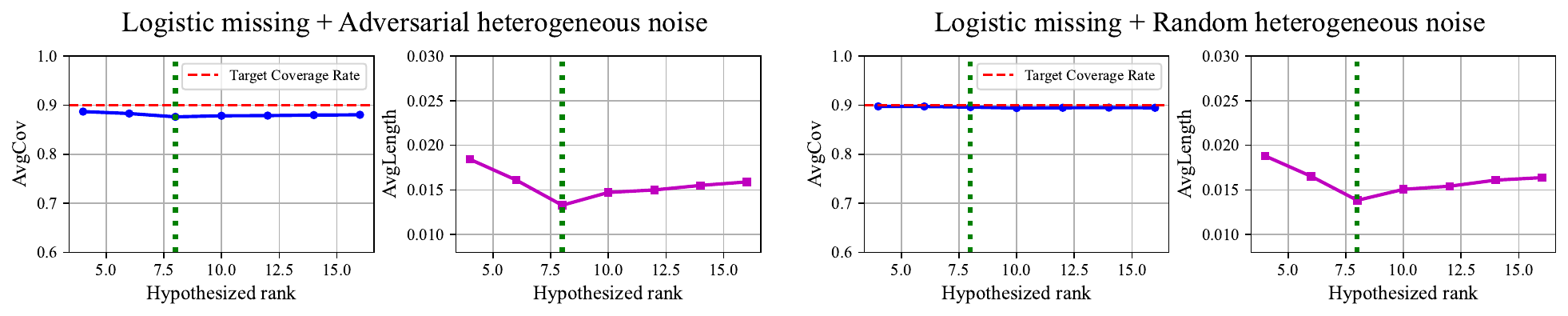}
	\caption{Comparison of recovery effects under heterogeneous noise}
	\label{FIG:3}
\end{figure}

\begin{figure}
	\centering
		\includegraphics[scale=0.45]{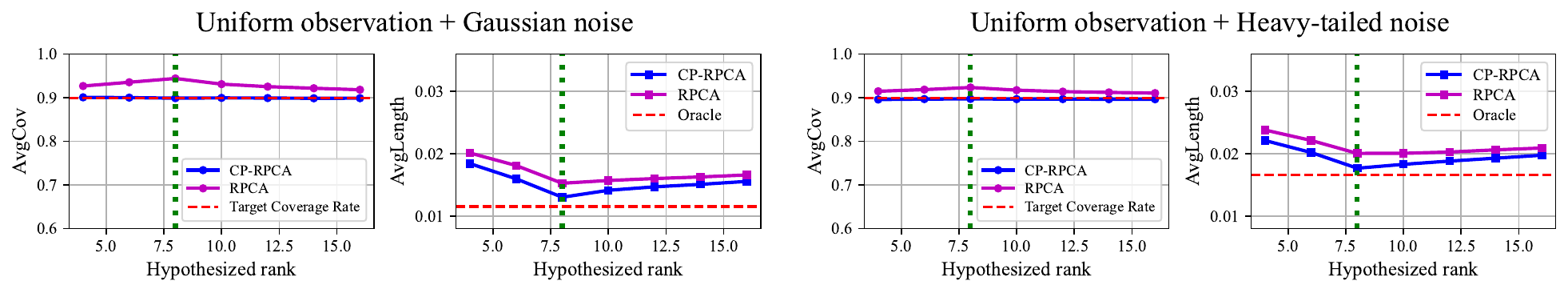}
	\caption{Comparison between RPCA and CP-RPCA}
	\label{FIG:4}
\end{figure}

Figure~\ref{FIG:3} reports the corresponding coverage rates and average interval lengths. Under random heterogeneous noise, the variability in the standardized residuals $\widehat{R}_{ij}$ is unstructured, resulting in only mild distributional shift and coverage rates close to the nominal level. In contrast, adversarial heterogeneous noise induces a more pronounced shift, as the noise variance is intentionally aligned with errors in estimating $\widehat{p}_{ij}$. In particular, high-variance noise is concentrated on entries with small observation probabilities, where the estimation error of $\widehat{p}_{ij}$ is typically largest. This adversarial alignment amplifies the impact of weighting errors on the conformal scores, leading to modest undercoverage. Nevertheless, the deviation from the target coverage remains limited, demonstrating the robustness of CP-RPCA even under adversarial heteroscedasticity. In Section~\ref{appendix_ex} of the Appendix, we further present the experimental results on the effects of sample size and sparsity, together with the results under misspecified observation models.  

\subsection{Comparison with Bootstrap-based RPCA inference}

Most existing RPCA methods are primarily designed for accurate point estimation under specific modeling assumptions, while systematic uncertainty quantification for the recovered low-rank component remains largely unexplored. In particular, to the best of our knowledge, there currently exist no established inference procedures for RPCA that provide entrywise confidence intervals and can serve as a direct benchmark for comparison with our proposed CP-RPCA method. Moreover, model-based asymptotic inference for RPCA—analogous to that developed for matrix completion in \citet{gui2023conformalized}—remains underdeveloped, making such comparisons infeasible at present. While developing asymptotic, model-driven inference guarantees for RPCA constitutes an important and appealing research direction in its own right, it lies beyond the scope of this work and is deferred to future investigation.

As a pragmatic baseline, we adopt a bootstrap-based RPCA inference procedure. Specifically, we estimate pointwise variances of the low-rank estimates via bootstrap resampling and construct confidence intervals by augmenting the point estimates with normal quantiles, following the strategy used in \citet{gui2023conformalized}. This approach serves as a natural reference for evaluating the empirical performance of CP-RPCA.

Figure~\ref{FIG:4} compares the resulting confidence intervals from the bootstrap-based RPCA method and CP-RPCA. The bootstrap-based approach exhibits substantial overcoverage, which can be attributed to its relatively poor reconstruction performance on unobserved entries. In particular, errors in low-rank recovery on the test set inflate the estimated variance beyond the true noise level, leading to overly conservative and excessively wide intervals. In contrast, CP-RPCA calibrates its prediction thresholds using a held-out calibration set and, through weighted exchangeability, effectively transfers distributional information from the calibration data to the test points. By leveraging this data-driven calibration mechanism, CP-RPCA produces confidence intervals that are substantially less conservative while maintaining valid coverage, thereby offering a more accurate and informative quantification of uncertainty.

\section{Real data application}

In this section, we demonstrate the practical utility of the proposed CP-RPCA method in image processing through two representative case studies.

\subsection{Feature extraction in face images}

It is well known that, under varying illumination conditions, images of a convex Lambertian surface approximately lie in a low-dimensional linear subspace \citep{basri2003lambertian}. This observation underpins the empirical success of low-rank models in face image analysis, where collections of images of the same individual can often be well approximated by a low-dimensional representation. In practice, however, real-world face images are frequently contaminated by shadows, specular highlights and  saturation effects. These artifacts are typically large in magnitude but spatially sparse, which can substantially degrade the quality of low-rank feature extraction and impair recognition performance. Our proposed RPCA can be used as an effective tool for separating such sparse corruptions from the underlying low-rank structure when a sufficient number of images of the same subject are available. Building on this principle, we apply the proposed CP-RPCA method to recover low-rank facial features while simultaneously constructing entrywise confidence intervals, thereby quantifying uncertainty in the recovered representations and offering additional insights for face recognition tasks.

Our experiments are conducted on the cropped YaleB face database \citep{lee2005acquiring}, which consists of near-frontal images of 38 subjects captured under 64 distinct lighting conditions. The images are preprocessed by cropping based on facial landmarks (e.g., eye corners) to remove background and irrelevant regions and  samples with extreme illumination or low contrast are excluded. For each subject, the vectorized face images are stacked into an observation matrix $Y$, with each row corresponding to one image. We set the estimated rank to $\widehat{r}=4$; potential misspecification in the rank is subsequently accommodated through the constructed confidence intervals.

\begin{figure}
	\centering
		\includegraphics[scale=0.45]{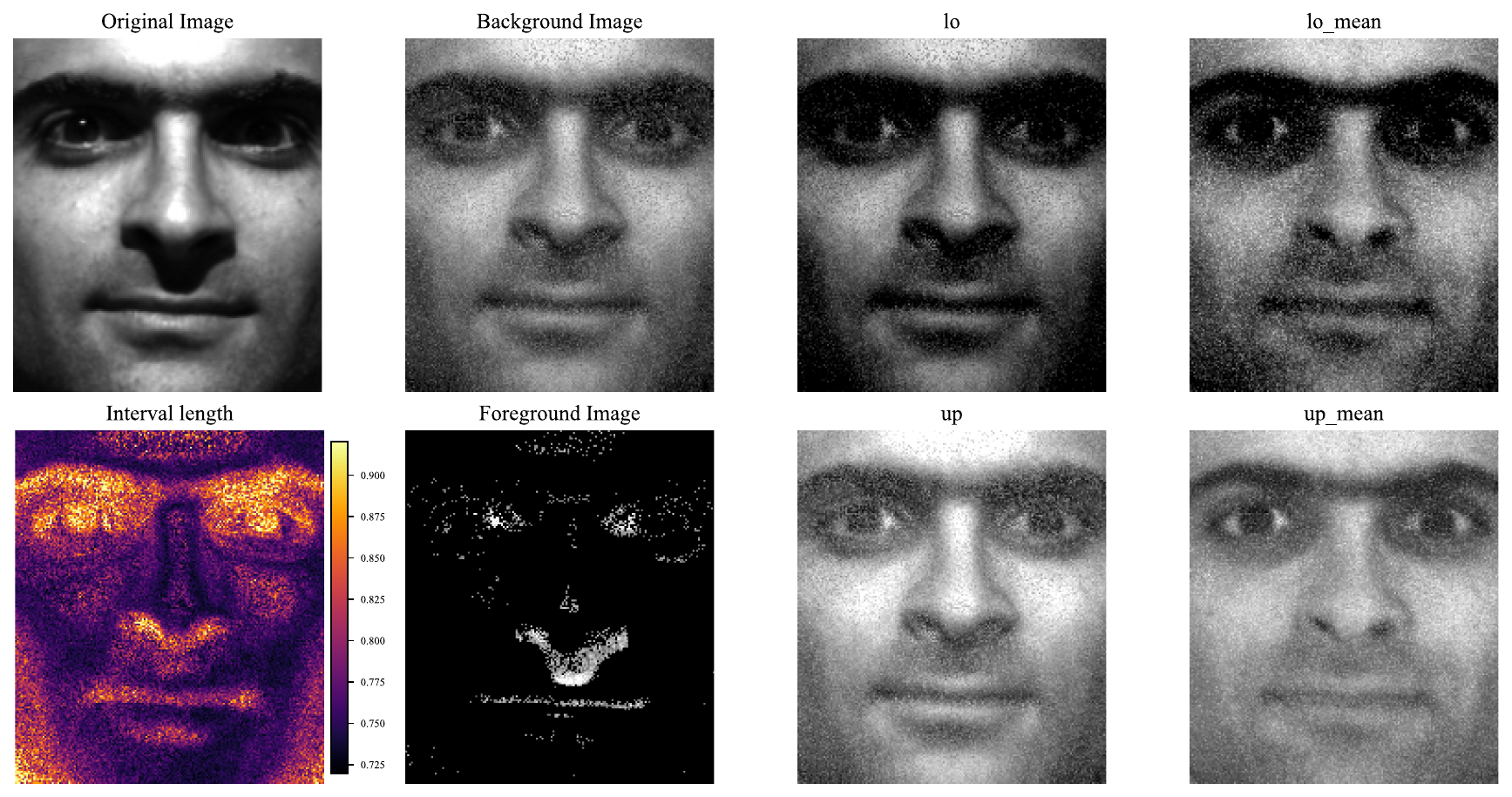}
	\caption{YaleB11 Face Feature Extraction and Confidence Interval}
	\label{FIG:5}
\end{figure}

Figure~\ref{FIG:5} illustrates the results for the YaleB11 subject. 
(1) The recovered low-rank component captures the stable facial structure, while the sparse component isolates illumination-induced variations around regions such as the forehead and nose, as well as expression-related changes around the eyes and mouth. 
(2) The lower and upper bounds of the confidence interval (denoted by ``lo'' and ``up'') for a single image account for uncertainty arising from rank estimation and model approximation, with approximately $90\%$ of the original pixel intensities falling within the interval. 
(3) The quantities ``lo\_mean'' and ``up\_mean'' represent the averaged interval bounds across multiple images of the same subject, yielding smoother boundaries and providing coverage guarantees across diverse lighting conditions. Notably, the resulting confidence intervals are wider in regions with higher variability, such as the eye corners and lips, reflecting greater uncertainty in the low-rank recovery in these areas. By examining the spatial distribution of interval widths, our proposed CP-RPCA enables the identification of regions where low-rank modeling is less reliable, suggesting opportunities for localized geometric modeling or adaptive feature extraction to further enhance face recognition performance.

\subsection{Video background modeling}\label{sec5.2}

Owing to strong temporal correlations across frames, the background component of video data is often well approximated by a low-rank structure. A fundamental task in video surveillance is background modeling, where the objective is to recover a static or slowly varying background while simultaneously detecting moving objects in the foreground \citep{candes2011robust, vaswani2018static}. In practice, this task is challenging for several reasons. First, in crowded or dynamic scenes, foreground objects may appear in nearly every frame, making it difficult to identify clean background observations. Second, background components themselves may evolve over time due to gradual illumination changes or environmental variations, requiring sufficient modeling flexibility. A common and effective modeling strategy is to treat the background as approximately low-rank, while foreground objects—such as pedestrians or vehicles—occupy only a small fraction of pixels in each frame and thus can be modeled as sparse components. Under this formulation, RPCA provides a principled approach for foreground–background separation. However, in real-world videos, the sparse component often exhibits spatial and temporal structure, for example in the form of contiguous moving regions. Such structured sparsity can violate the idealized assumptions of RPCA and lead to imperfect recovery of the low-rank background.

In this section, we apply our proposed CP-RPCA framework  to video background modeling with a particular emphasis on uncertainty quantification. Beyond point estimation, CP-RPCA constructs entrywise confidence intervals for the recovered background, thereby compensating for residual errors arising from model misspecification (e.g., incorrect rank specification; see \citealt{xu2024lere}) or structured sparse noise. Importantly, even when the quality of background recovery is imperfect, the CP–based intervals continue to provide reliable coverage guarantees, highlighting the robustness of the proposed approach in realistic settings.

\begin{figure}
	\centering
		\includegraphics[scale=0.48]{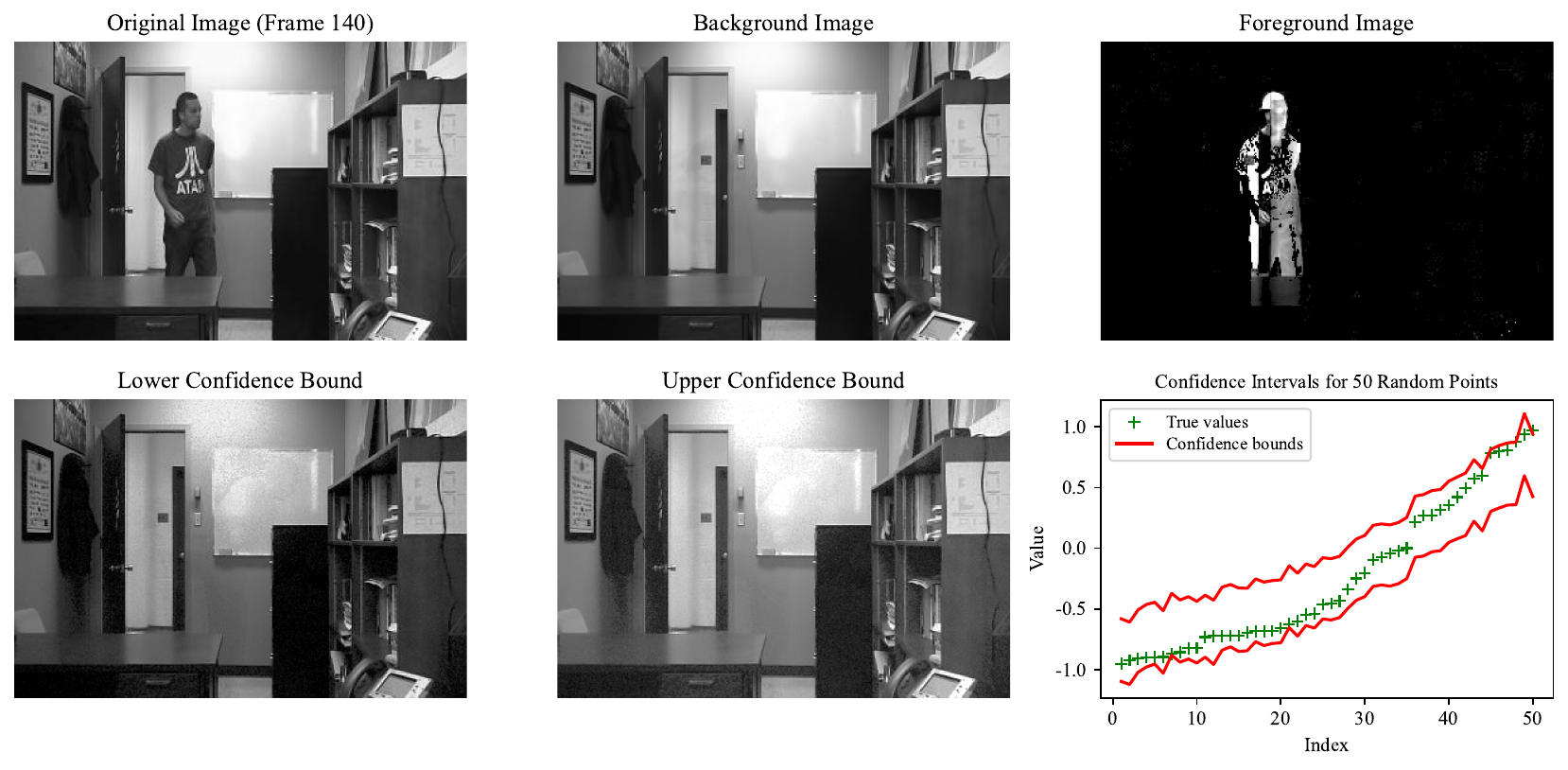}
	\caption{Video foreground-background separation results and confidence intervals}
	\label{FIG:6}
\end{figure}

We consider a representative surveillance video from the CDnet dataset \citep{wang2014cdnet}, which features a largely static background with substantial foreground activity. A total of 200 frames are extracted, each with resolution $240\times 360$. Each frame is vectorized and stacked as a row, yielding an observation matrix $Y\in\mathbb{R}^{200\times 86{,}400}$. To evaluate coverage, we randomly select $5\%$ of the entries as a test set, which are excluded from both model fitting and calibration. The goal is to decompose $Y$ into a low-rank background component $X$ and a sparse foreground component $S$ and  to construct confidence intervals $C(X)$ for the background with a target coverage level of $90\%$.

An inspection of the singular value decay reveals a clear elbow, with the first two singular values dominating the spectrum. Accordingly, we set the assumed rank to $r=2$. Figure~\ref{FIG:6} presents the results for the $140$th frame as a representative example. The recovered background captures the static scene structure, while moving objects are effectively isolated in the sparse component. Moreover, the constructed confidence intervals for the background are relatively narrow and achieve near-nominal coverage when evaluated on 50 randomly selected test pixels. These results demonstrate that CP-RPCA can provide both interpretable background reconstructions and reliable uncertainty quantification in video data, even in the presence of structured foreground motion.

\section{Discussions}


This paper introduces CP-RPCA, a novel framework that integrates CP with RPCA. By leveraging the distribution-free nature of CP, our method provides entrywise confidence intervals for matrix recovery without restrictive distributional assumptions. We establish non-asymptotic coverage guarantees, proving that the interval lengths are adaptive to model misspecification and approach optimality under correct specifications. Both numerical simulations and empirical applications validate the robustness and practical utility of CP-RPCA. Despite the strengths of CP-RPCA, several avenues for future research remain: First, the split-conformal implementation inevitably introduces additional randomness due to data splitting. Developing more stable procedures that aggregate information across multiple splits—such as cross-conformal or cross-validation–based approaches—would therefore be practically appealing. Existing frameworks proposed by \citet{lei2018distribution} and \citet{lei2020cross} provide a natural and promising starting point for such extensions. Second, while our theoretical guarantees are finite-sample and distribution-free, it is of interest to study asymptotic coverage behavior and to characterize the coverage gap as the sample size grows. Prior work has investigated asymptotic coverage in regression settings under exchangeability \citep{lei2018distribution, angelopoulos2024theoretical}, often for specific scoring rules such as the conformalized quantile regression (CQR) score \citep{sesia2020comparison}. Extending such results to structured matrix problems remains an open challenge. Third, achieving approximate conditional coverage with finite-length intervals is an important yet fundamentally difficult goal. Recent studies have explored relaxed notions of conditional validity, including test-conditional coverage \citep{barber2021predictive}, label-conditional coverage via clustering \citep{ding2023class} and  matrix-specific notions such as row- and row–column–conditional validity \citep{shao2023distribution}. These ideas offer useful insights for extending conformal inference to richer notions of validity in matrix-structured settings.


Beyond RPCA, we aim to extend the CP-RPCA framework to a broader class of low-rank recovery problems, including 1-bit matrix completion, principal component analysis and  online matrix completion. In 1-bit matrix completion, only the signs of the underlying entries are observed, resulting in severe information loss. \cite{chen2023statistical} studied statistical inference in this setting under explicit probabilistic models, establishing minimax-optimal rates and asymptotically valid inference procedures. Extending distribution-free uncertainty quantification to such highly quantized data remains an important and challenging open problem. For principal component analysis, \cite{yan2024inference} developed inference theory for eigenvalues and eigenvectors under heteroskedastic noise and non-uniform missingness, deriving nonasymptotic guarantees that critically rely on accurate modeling of both noise variances and the observation mechanism. Their results highlight the sensitivity of classical, model-based PCA inference to nuisance parameter misspecification, further motivating robust, distribution-free alternatives. In online matrix completion, observations arrive sequentially, posing fundamental challenges to batch-based conformal methods. Our goal is to develop incremental CP procedures that update low-rank estimates and confidence intervals in a streaming fashion without repeated refitting. Prior work has emphasized the importance of this problem in dynamic recommendation and decision-making systems \citep{pal2022online, baby2024online}. More recently, \cite{han2025online}, \cite{duan2025online} and \cite{duan2025statistical} advanced model-based inference for online matrix completion under adaptive and dependent observation mechanisms, developing nonconvex estimation and debiasing techniques. While these methods provide sharp guarantees under carefully specified models, they also underscore the need for uncertainty quantification tools that remain valid under weaker assumptions—an opportunity well suited to CP.


\bibliographystyle{cas-model2-names}

\bibliography{cas-refs}

\appendix

\section{Underlying RPCA Solver: Fast-RPCA}\label{appendix_01}

To obtain accurate point estimates of the low-rank component $X$, we adopt the Fast-RPCA algorithm proposed by \citet{yi2016fast}, together with a reliable initialization scheme inspired by \citet{zhang2018unified}. This algorithm serves as the underlying RPCA solver throughout our experiments.

\paragraph{Nonconvex formulation.}
Unlike convex relaxation–based RPCA methods, Fast-RPCA employs a nonconvex factorized representation of the low-rank matrix $X = UV^\top$, where $U \in \mathbb{R}^{d_1 \times r}$ and $V \in \mathbb{R}^{d_2 \times r}$. The recovery problem is formulated as
\begin{equation}\label{eq06}
    \min_{U,V,S} \mathcal{L}_n (UV^\top+S), \quad \text{subject to} \quad U\in \mathcal{C}_1, V \in \mathcal{C}_2, \|S\|_0 \le s,
\end{equation}
where $\mathcal{L}_n(\cdot)$ denotes the empirical loss under partial observation, $\mathcal{C}_1$ and $\mathcal{C}_2$ are constraint sets enforcing incoherence, and $s$ is the sparsity level of the corruption matrix $S^\ast$. This factorized formulation yields nearly linear computational complexity in the ambient dimensions when $r \ll \min(d_1,d_2)$, while retaining strong robustness guarantees.

\paragraph{Structural assumptions.}
To ensure identifiability of the low-rank and sparse components, the following assumptions are imposed.
(1) \textbf{Incoherence of the low-rank component.}
Let $X = U^\ast \Sigma^\ast V^{\ast\top}$ be the SVD of $X$. We assume $\mu$-incoherence:
$$
\|U^\ast\|_{2,\infty} \le \sqrt{\mu r / d_1}, \qquad \|V^\ast\|_{2,\infty} \le \sqrt{\mu r / d_2},
$$
where $\mu \ge 1$ represents the incoherence parameter.	
(2) \textbf{Dispersed sparsity of the corruption.}
The sparse component $S^\ast$ is assumed to have at most a fraction $\beta \in (0, 1)$ of nonzero entries in each row and column.   Denote $\text{S}^*\in \mathcal{K} $, where $ \mathcal{K} $ is defined as follows:
$$
\mathcal{K} = \{S\in \mathbb{R}^{d_1 \times d_2}: \|S\|_0 \le s, \|S_{i,\cdot}\|_0 \le \beta d_2, \|S_{\cdot,j}\|_0 \le \beta d_1\}.
$$

\paragraph{Constraint sets.}
Under these assumptions, the constraint sets in \eqref{eq06} are specified as
\begin{equation}\label{eq07}
    \mathcal{C}_1 = \{U: \|U\|_{2,\infty} \le \sqrt{\mu r/d_1}\|G^0\|_2\}, \quad \mathcal{C}_2 = \{V: \|V\|_{2,\infty} \le \sqrt{\mu r/d_2} \|G^0\|_2\},
\end{equation}
where $G^0 = [U^0; V^0]$ denotes the initial factor estimate.

\paragraph{Optimization objective.}
Fast-RPCA minimizes the regularized objective
$$
\min_{U,V,S} \mathcal{L}(U,V;S) + \frac{1}{64} \|U^\top U - V^\top V\|_F^2, \quad \text{subject to} U \in \mathcal{C}_1, V \in \mathcal{C}_2, S \in \mathcal{K},
$$
where the additional penalty term enforces scale balance between $U$ and $V$.
where $\mathcal{L}(U,V;S)$ denotes the empirical loss function under partial observations and  the term $\tfrac{1}{64}\|U^{\top}U-V^{\top}V\|_F^2$ serves as an additional regularizer that penalizes scale imbalance between $U$ and $V$. The constraint sets $\mathcal{C}_1$ and $\mathcal{C}_2$ are defined in \eqref{eq07}.

\paragraph{Gradient-based algorithm.}
The optimization is solved via projected gradient descent. The full procedure is summarized in Algorithm~\ref{alg:fast_rpca}.

\begin{algorithm}[ht]
\caption{\textbf{Fast-RPCA under partial observation}}\label{alg:fast_rpca}

\textbf{Inputs:} Observed matrix $Y$ with support $\mathcal{D}$, step size $\eta$, number of iterations $T$, threshold parameter $\gamma$, initialization $(U^0,V^0,S^0)$.

\begin{algorithmic}[1]

\State $G^{0} = [U^{0}; V^{0}]$; \quad Calculate $\mathcal{C}_{1}, \mathcal{C}_{2}$ according to Eq.~\eqref{eq07}.

\For{$t = 0,1,\dots, T-1$}

\State $S^{t} = \mathcal{T}_{\gamma p \alpha}[\Pi_{\mathcal{D}} (Y - U^{t} V^{t\top})]$

\State $U^{t+1} = \Pi_{\mathcal{C}_{1}}(U^{t} - \eta \nabla_{U} \mathcal{L}(U^{t}, V^{t}; S^{t}) - \eta U^{t}(U^{t\top} U^{t} - V^{t\top} V^{t})/16)$

\State $V^{t+1} = \Pi_{\mathcal{C}_{2}}(V^{t} - \eta \nabla_{V} \mathcal{L}(U^{t}, V^{t}; S^{t}) - \eta V^{t}(V^{t\top} V^{t} - U^{t\top} U^{t})/16)$

\EndFor

\end{algorithmic}

\textbf{Outputs:} $(U^{T}, V^{T}, S^{T})$.

\end{algorithm}

\noindent Here $\Pi_{\mathcal{C}_i}$ denotes projection onto $\mathcal{C}_i$, $i=1,2$, and $\mathcal{T}_{\beta}: \mathbb{R}^{d_1\times d_2} \rightarrow \mathbb{R}^{d_1\times d_2}$ is the truncation operator. For all $(i,j) \in [d_1] \times [d_2]$, we define
$$
[\mathcal{T}_{\beta}(S)]_{ij} =
\begin{cases}
    S_{ij}, 
    & \text{if }
    |S_{ij}| \ge |S_{(i,\cdot)}|^{(\beta d_2)}
    \ \text{and}\
    |S_{ij}| \ge |S_{(\cdot,j)}|^{(\beta d_1)}, \\
    0, 
    & \text{otherwise}.
\end{cases}
$$
where $|S_{(i,\cdot)}|^{(k)}$ and $|S_{(\cdot,j)}|^{(k)}$ represent the $k$-th largest absolute values in $S_{(i,\cdot)}$ and $S_{(\cdot,j)}$, respectively. Through this operator, we retain the entries in the top $\beta$-quantile of absolute values in both the corresponding row and column.

\paragraph{Initialization.}
\cite{zhang2018unified} showed that, provided the initial solution $(U^0, V^0, S^0)$ lies in a sufficiently small neighborhood of the true solution $(U^\ast, V^\ast, S^\ast)$, Algorithm~\ref{alg:fast_rpca} converges linearly to the global optimum. Motivated by this result, we adopt the initialization strategy proposed in \cite{zhang2018unified} to ensure a reliable starting point for the Fast-RPCA algorithm. Specifically, the initialization procedure combines hard thresholding and low-rank projection. We first apply a hard thresholding operator $\mathcal{H}_k$, which retains the $k$ entries with the largest absolute values in a matrix and sets all remaining entries to zero, to obtain an initial estimate of the sparse component. To initialize the low-rank component, we further introduce a constrained projection operator $\mathcal{P}_{\zeta^\ast}$ that integrates singular value projection with an additional entrywise infinity-norm constraint, thereby preventing spiky low-rank estimates. Formally, $\mathcal{P}_{\zeta^\ast}$ is defined as
$$
\mathcal{P}_{\zeta^\ast}(X) = \arg\min_{\substack{\operatorname{rank}(L) \le r \\ \|L\|_{\infty,\infty} \le \zeta^\ast}} \|L - X\|_F,
$$
where $\zeta^\ast = c_0 \mu r \kappa / \sqrt{d_1 d_2}$ and $c_0$ is a predetermined upper bound on the largest singular value $\sigma_1(X)$. In practice, this constrained projection can be efficiently implemented via the Dykstra alternating projection algorithm. Empirically, a single iteration of alternating projection is sufficient to produce an initialization that satisfies the conditions required for fast convergence. The complete initialization procedure is summarized in Algorithm~~\ref{alg:init} below.

\begin{algorithm}
\caption{Initialization Phase}\label{alg:init}
\begin{algorithmic}[1]
\Require step size $\tau, \eta$, number of iterations $L$, parameters $\lambda$.
\State $X_0 = S_0 = 0$
\For{$l = 0, 1, \cdots, L-1$}
    \State $S_{l+1} = \mathcal{H}_{\lambda s}(S_l - \tau \nabla_S \mathcal{L}_n(X_l + S_l))$
    \State $X_{l+1} = \mathcal{P}_{\zeta^\ast}( X_l - \eta \nabla_X \mathcal{L}_n(X_l + S_l))$
\EndFor
\State $[\bar{U}^0, \Sigma^0, \bar{V}^0] = \mathrm{SVD}_r(X_L)$
\State $U^0 = \bar{U}^0 (\Sigma^0)^{1/2}, V^0 = \bar{V}^0 (\Sigma^0)^{1/2}, S^0 = S_L$
\Ensure $(U^0, V^0, S^0)$
\end{algorithmic}
\end{algorithm}

\section{Theoretical Guarantees and Proofs}

\subsection{Proof of Lemma~\ref{lem01}}

We begin by reviewing the CP-RPCA data-generating and sampling mechanism. The observation set $\Omega_{\mathrm{obs}}$  is generated according to the observation probability matrix $P$ and the support matrix $Z$. It is randomly split into a calibration set $\Omega_{\mathrm{cal}}$ and a training set $\Omega_{\mathrm{tr}}$ with proportion $q$, so that
$$
\Omega_{\mathrm{cal}}\cup \Omega_{\mathrm{tr}}=\Omega_{\mathrm{obs}}.
$$
Each element in $\Omega_{\mathrm{cal}}$ is independently contaminated with probability $\beta$. After removing contaminated entries, we obtain the effective calibration set $\Omega_{\mathrm{cal}}^{\prime}$ and the discard set $\Omega_{\mathrm{drop}}$, satisfying
$$
\Omega_{\mathrm{cal}}^{\prime}\cup \Omega_{\mathrm{drop}}=\Omega_{\mathrm{cal}}.
$$
The conformal weights are computed on $\Omega_{\mathrm{cal}}^{\prime}$ and subsequently normalized.
Fix a candidate set of locations
$$ 
\mathcal{B}=\{(i_1,j_1),\cdots,(i_{n_{\mathrm{cal}}^{\prime}+1},j_{n_{\mathrm{cal}}^{\prime}+1})\},  
$$
and an index $m\in \{1,\cdots,n_{\mathrm{cal}}^{\prime}+1\}$. By the definition of conditional probability,
\begin{eqnarray}\label{proof_eq01}
  & & \mathbb{P}\big((i_\ast,j_\ast)=(i_m,j_m) \mid \Omega_{\mathrm{cal}}^{\prime} \cup \{(i_\ast,j_\ast)\} = \mathcal{B}, \Omega_{\mathrm{tr}} = \Omega_0, \Omega_{\mathrm{drop}} = \Omega_2\big) \nonumber\\
  &=& \frac{\mathbb{P}\big((i_\ast,j_\ast) = (i_m,j_m), \Omega_{\mathrm{cal}}^{\prime} = \mathcal{B} \setminus \{(i_m,j_m)\} \mid \Omega_{\mathrm{tr}} = \Omega_0, \Omega_{\mathrm{drop}} = \Omega_2,  |\Omega_{\mathrm{obs}}| = n\big)}{
\sum_{l=1}^{n_{\mathrm{cal}}^{\prime}+1} \mathbb{P}\big((i_\ast,j_\ast) = (i_l,j_l), \Omega_{\mathrm{cal}}^{\prime} = \mathcal{B} \setminus \{(i_l,j_l)\} \mid \Omega_{\mathrm{tr}} = \Omega_0, \Omega_{\mathrm{drop}} = \Omega_2, |\Omega_{\mathrm{obs}}| = n\big)}. 
\end{eqnarray}
For any fixed set $\Omega_1$ and any location $(i_l,j_l) \notin \Omega_{\mathrm{obs}}$, consider
\begin{eqnarray}\label{proof_eq02}
    & & \mathbb{P}\big(\Omega_\mathrm{cal}^{\prime}=\Omega_1, (i_\ast,j_\ast)=(i_l,j_l) \mid \Omega_{\mathrm{tr}}=\Omega_0, \Omega_{\mathrm{drop}}=\Omega_2, |\Omega_{\mathrm{obs}}|=n\big) \nonumber\\
    &=& \frac{\mathbb{P}\big(\Omega_\mathrm{cal}^{\prime}=\Omega_1, \Omega_{\mathrm{tr}}=\Omega_0, \Omega_{\mathrm{drop}}=\Omega_2, (i_\ast,j_\ast)=(i_l,j_l) \mid |\Omega_{\mathrm{obs}}|=n\big)}{\mathbb{P}\big(\Omega_{\mathrm{tr}}=\Omega_0, \Omega_{\mathrm{drop}}=\Omega_2 \mid |\Omega_{\mathrm{obs}}|=n\big)} \nonumber\\
    &=& \frac{\mathbb{P}\big((i_\ast,j_\ast)=(i_l,j_l) \mid \Omega_{\mathrm{obs}}=\Omega_1\cup\Omega_0\cup\Omega_2\big) \cdot \mathbb{P}\big(\Omega_\mathrm{cal}^{\prime}=\Omega_1, \Omega_{\mathrm{tr}}=\Omega_0, \Omega_{\mathrm{drop}}=\Omega_2 \mid |\Omega_{\mathrm{obs}}|=n\big)}{\mathbb{P}\big(\Omega_{\mathrm{tr}}=\Omega_0, \Omega_{\mathrm{drop}}=\Omega_2 \mid |\Omega_{\mathrm{obs}}|=n\big)} \nonumber\\
    &=& \frac{1}{d_1 d_2 - n}\cdot\frac{\mathbb{P}\big(\Omega_\mathrm{cal}^{\prime}=\Omega_1, \Omega_{tr}=\Omega_0, \Omega_{\mathrm{drop}}=\Omega_2 \mid |\Omega_{\mathrm{obs}}|=n\big)}{\mathbb{P}\big(\Omega_{\mathrm{tr}}=\Omega_0, \Omega_{\mathrm{drop}}=\Omega_2 \mid |\Omega_{\mathrm{obs}}|=n\big)}.  
\end{eqnarray}
Since
\begin{eqnarray*}
    & & \mathbb{P}\big(\Omega_\mathrm{cal}=\Omega_3, \Omega_{\mathrm{tr}}=\Omega_0 \mid |\Omega_{\mathrm{obs}}|=n\big) \\
    &=& \mathbb{P}\big(\operatorname{supp}(Z)=\Omega_0\cup\Omega_3, \Omega_0\subseteq\operatorname{supp}(W), \Omega_3\subseteq\operatorname{supp}(W)^c \mid |\Omega_{\mathrm{obs}}|=n\big) \\
    &=& \mathbb{P}\big(\Omega_0\subseteq\operatorname{supp}(W), \Omega_3\subseteq\operatorname{supp}(W)^c\big) \cdot \mathbb{P}\big(\operatorname{supp}(Z)=\Omega_0\cup\Omega_3 \mid |\Omega_{\mathrm{obs}}|=n\big) \\
    &=& \frac{q^{|\Omega_0|}(1-q)^{|\Omega_3|}}{\mathbb{P}(|\Omega_{\mathrm{obs}}|=n)}\prod_{(i^{\prime},j^{\prime})\in[d_1]\times[d_2]} (1-p_{i^{\prime}j^{\prime}})\prod_{(i,j)\in\Omega_{\mathrm{obs}}} \frac{p_{ij}}{1-p_{ij}}. 
\end{eqnarray*}
Then, we have
\begin{eqnarray}\label{proof_eq03}
    & & \mathbb{P}\big(\Omega_\mathrm{cal}^{\prime}=\Omega_1, \Omega_{\mathrm{tr}}=\Omega_0, \Omega_{\mathrm{drop}}=\Omega_2 \mid |\Omega_{\mathrm{obs}}|=n\big) \nonumber\\
    &=& \mathbb{P}\big(\Omega_\mathrm{cal}=\Omega_3, \Omega_{\mathrm{tr}}=\Omega_0 \mid |\Omega_{\mathrm{obs}}|=n\big) \cdot \mathbb{P}\big(\Omega_\mathrm{cal}^{\prime}=\Omega_1, \Omega_{\mathrm{drop}}=\Omega_2 \mid \Omega_{\mathrm{tr}}=\Omega_0, \Omega_3 = \Omega_1\cup\Omega_2, |\Omega_{\mathrm{obs}}|=n\big) \nonumber\\ 
    &=& \frac{q^{|\Omega_0|}(1-q)^{|\Omega_3|}\beta^{|\Omega_2|}(1-\beta)^{|\Omega_1|}}{\mathbb{P}(|\Omega_{\mathrm{obs}}|=n)}\prod_{(i^{\prime},j^{\prime})\in [d_1]\times[d_2]} (1-p_{i^{\prime}j^{\prime}})\prod_{(i,j)\in \Omega_{\mathrm{obs}}} \frac{p_{ij}}{1-p_{ij}}.
\end{eqnarray}
Combining this with the joint probability of the splitting and contamination mechanisms, we obtain
\begin{eqnarray}\label{proof_eq04}
    & & \mathbb{P}(\Omega_{\mathrm{tr}}=\Omega_0,\Omega_{\mathrm{drop}}=\Omega_2 \mid |\Omega_{\mathrm{obs}}|=n) \nonumber\\
    &=& \sum_{\mathcal{A}\in\Omega}\mathbb{P}(\Omega_{\mathrm{cal}}^{\prime}=\mathcal{A},\Omega_{\mathrm{tr}}=\Omega_0,\Omega_{\mathrm{drop}}=\Omega_2 \mid |\Omega_{\mathrm{obs}}|=n) \nonumber\\
    &=& \frac{q^{|\Omega_0|}(1-q)^{|\Omega_3|}\beta^{|\Omega_2|}(1-\beta)^{|\Omega_1|}}{ \mathbb{P}(|\Omega_{\mathrm{obs}}|=n)}\sum_{\mathcal{A}\in\Omega}\Big(\prod_{(i^{\prime},j^{\prime})\in[d_1]\times[d_2]} (1-p_{i^{\prime}j^{\prime}})\prod_{(i,j)\in\Omega_{\mathrm{obs}}} \frac{p_{ij}}{1-p_{ij}}\Big), 
\end{eqnarray}
where
\begin{equation*}
    \Omega =\big\{\mathcal{A}\subseteq [d_1] \times [d_2]: |\mathcal{A}|=n-|\Omega_{\mathrm{tr}}|-|\Omega_{\mathrm{drop}}|,\mathcal{A}\cap \Omega_{\mathrm{tr}}=\varnothing, \mathcal{A}\cap \Omega_{\mathrm{drop}}=\varnothing\big\}. 
\end{equation*}
Substituting \eqref{proof_eq03} and \eqref{proof_eq04} into \eqref{proof_eq02} yields
\begin{eqnarray}\label{proof_eq05}
    & & \mathbb{P}\big(\Omega_\mathrm{cal}^{\prime}=\Omega_1, (i_\ast,j_\ast)=(i_l,j_l) \mid \Omega_{\mathrm{tr}}=\Omega_0, \Omega_{\mathrm{drop}}=\Omega_2, |\Omega_{\mathrm{obs}}|=n\big) \nonumber\\
    &=& \frac{1}{d_1d_2 - n} \frac{\prod_{(i,j)\in \Omega_1}\frac{p_{ij}}{1-p_{ij}}}{\sum_{\mathcal{A}\in \Omega}\prod_{(i^{\prime},j^{\prime})\in \mathcal{A}}\frac{p_{i^{\prime}j^{\prime}}}{1-p_{i^{\prime}j^{\prime}}}}.  
\end{eqnarray}
Substituting \eqref{proof_eq05} into the conditional probability in \eqref{proof_eq01} further gives
\begin{eqnarray*}
    & & \mathbb{P}\big((i_\ast,j_\ast)=(i_m,j_m) \mid \Omega_{\mathrm{cal}}^{\prime} \cup \{(i_\ast,j_\ast)\} = \mathcal{B}, \Omega_{\mathrm{tr}} = \Omega_0, \Omega_{\mathrm{drop}} = \Omega_2\big) \\
    &=& \frac{\prod_{(i,j)\in \mathcal{B}\setminus \{(i_m,j_m)\}}\frac{p_{ij}}{1-p_{ij}}}{\sum_{l=1}^{n_{\mathrm{cal}}^{\prime}+1}\prod_{(i,j)\in \mathcal{B}\setminus \{(i_l,j_l)\}}\frac{p_{ij}}{1-p_{ij}}} = \frac{h_{i_m j_m} }{\sum_{l=1}^{n_{\mathrm{cal}}^{\prime}+1} h_{i_lj_l}}. 
\end{eqnarray*}

\subsection{Proof of Theorem~\ref{th01}}

To prove Theorem~\ref{th01}, we first establish two auxiliary lemmas concerning weighted split CP under correctly specified observation probabilities.

\begin{lemma}\label{lem03}
    Suppose the observation probability matrix $P$ is known and the weights are correctly specified. For any symmetric score function $R$, the confidence interval $\widehat{C}$ produced by split CP for an observed location $(i_{\ast},j_{\ast})\mid\Omega_{\mathrm{obs}}\sim\text{Unif}(\Omega_{\mathrm{obs}}^c)$ satisfies
    \begin{equation*}
    \mathbb{P}\big\{X_{i_\ast j_\ast} \in \widehat{C}(i_\ast j_\ast)\big\} = \mathbb{P}\big\{X_{i_\ast j_\ast} \in \widehat{X}_{i_\ast j_\ast} \pm q_{i_\ast j_\ast}^{\ast} \cdot \widehat{\sigma}_{i_\ast j_\ast}\big\} = \mathbb{P}\big\{R_{i_\ast j_\ast} \le q_{ii_\ast j_\ast}^{\ast}\big\} \ge 1 - \alpha,
    \end{equation*}
    where  $q_{i_{\ast}j_{\ast}}^{\ast}=\operatorname{Quantile}_{1-\alpha}\big(\sum_{k=1}^{n_{\mathrm{cal}}}{\omega_{i_kj_k}\delta_{R_{i_kj_k}}+\omega_{i_{\ast}j_{\ast}}\delta_{+\infty}}\big)$ is the weighted $(1-\alpha)$-quantile.
\end{lemma}

\begin{lemma}\label{lem04}
    Under the same conditions as Lemma~\ref{lem03}, assume further that the score function $R$ has a continuous distribution. Then the split conformal confidence interval $\widehat{C}$ satisfies
    \begin{equation*}
        \mathbb{P}\big\{X_{i_\ast j_\ast} \in \widehat{C}(i_\ast j_\ast)\big\} = \mathbb{P}\big\{X_{i_\ast j_\ast} \in \widehat{X}_{i_\ast j_\ast} \pm q_{i_\ast j_\ast}^{\ast} \cdot \widehat{\sigma}_{i_\ast,j_\ast}\big\} = \mathbb{P}\big\{R_{i_\ast j_\ast} \le q_{i_\ast j_\ast}^{\ast}\big\} \le 1 - \alpha + 2\omega_{\max},
    \end{equation*}
    where $q_{i_{\ast}j_{\ast}}^{\ast}$ is defined as in Lemma~\ref{lem03} and $\omega_{\max}=\max_k \{\omega_{i_kj_k}\}$.
\end{lemma}

\paragraph{Proof of Theorem~\ref{th01}.} Based on Lemma~\ref{lem03}, we have
\begin{equation*}
    \mathbb{P}\big(X_{i_{\ast} j_{\ast}}\in \widehat{X}_{i_{\ast} j_{\ast}}\pm q_{i_{\ast}j_{\ast}}^{\ast}(\alpha +\Delta+\varepsilon) \cdot \widehat{\sigma}_{i_{\ast} j_{\ast}}\big) \ge 1-\alpha -\mathbb{E}[\Delta]-\mathbb{E}[\xi]. 
\end{equation*}
It therefore suffices to show that 
$$
\widehat{q}\ge q_{i_{\ast}j_{\ast}}^{\ast}(\alpha +\Delta +\xi),
$$ 
which follows if
\begin{equation*}
\operatorname{Quantile}_{1-\alpha}\Big(\sum_{(i,j)\in \Omega_{\mathrm{cal}}^{\prime}}\widehat{\omega}_{ij}\delta_{\widehat{R}_{ij}}\Big) \ge
\operatorname{Quantile}_{1-\alpha-\Delta-\xi}\Big(\sum_{(i,j)\in \Omega_{\mathrm{cal}}^{\prime}}\omega_{ij}\delta_{R_{ij}}\Big).
\end{equation*}
From the definitions in \eqref{eq04}–\eqref{eq05}, the total variation distances satisfy
\begin{equation*}
d_{\mathrm{TV}}\Big(\sum_{(i,j)\in\Omega_{\mathrm{cal}}^{\prime}}\widehat{\omega}_{ij}\delta_{R_{ij}}, \sum_{(i,j)\in\Omega_{\mathrm{cal}}^{\prime}}\omega_{ij}\delta_{R_{ij}}\Big) \le \frac{1}{2} \sum_{(i,j)\in\Omega_{\mathrm{cal}}^{\prime}\cup{(i\ast,j_\ast)}}|\widehat{\omega}_{ij}-\omega_{ij}| =: \Delta,
\end{equation*}
and
\begin{equation*}
d_{\mathrm{TV}}\Big(\sum_{(i,j)\in\Omega_{\mathrm{cal}}^{\prime}}\widehat{\omega}_{ij}\delta_{\widehat{R}_{ij}}, \sum_{(i,j)\in\Omega_{\mathrm{cal}}^{\prime}}\widehat{\omega}_{ij}\delta_{R_{ij}}\Big)\le\sum_{{\widehat{R}_{ij}\neq R_{ij}}}\widehat{\omega}_{ij} =: \xi,
\end{equation*}
where $d_{\mathrm{TV}}$ denotes the total variation distance. This establishes the desired lower bound on the coverage probability. Similarly, by Lemma~\ref{lem04} and the corresponding upper bound implied by the total variation distance, we obtain
$$
\widehat{q}\le q_{i_{\ast}j_{\ast}}^{\ast}(\alpha-\Delta-\xi),
$$
which in turn yields the stated upper bound on the coverage rate.

\paragraph{Proofs of Lemmas~\ref{lem03} and~\ref{lem04}.} Lemma~\ref{lem03} follows directly from the weighted exchangeability argument of \cite{tibshirani2019conformal}. We now provide the proof of Lemma~\ref{lem04}. By weighted exchangeability,
\begin{equation*}
X_{i_\ast,j_\ast}\in C(i_\ast,j_\ast) \iff R_{n+1} \le \operatorname{Quantile}_{1-\alpha}\Big(\sum_{k=1}^{n}\omega_k\delta_{R_k} + \omega_{n+1}\delta_{+\infty}\Big),
\end{equation*}
where $n=n_{cal}$, $R_{n+1}$ is the test score and  $R_{(1)}\le \cdots \le R_{(n)}<R_{+\infty}$ are the ordered scores. Define the weighted empirical CDF
\begin{equation*}
     F_{P,R}(t) =\sum_{i=1}^{n+1}{\omega_i\mathbb{1}\{R_i\le t\}}.
\end{equation*}
Let
$$ 
R_{(k)}=\sup\{t: \widehat{F}(t) \le 1-\alpha\}.
$$
Then non-coverage occurs if and only if $R_{n+1}\ge R_{(k)}$. By order-statistic arguments,
\begin{equation*}
     R_{n+1}\ge R_{(k)} \iff  \text{either}~~R_{n+1}>R_{(k-1)}~~\text{or}~~R_{n+1}=R_{(k)}=R_{(k-1)}.
\end{equation*}
If $R_{n+1}=R_{(k)}=R_{(k-1)}$, then for some $j\in [n]$, we have $R_{n+1}=R_j$ and  thus,
\begin{equation*}
\mathbb{P}\big(X_{i_{\ast}j_{\ast}}\notin C(i_{\ast},j_{\ast})\big) = \mathbb{P}\big(R_{n+1}>R_{(k-1)}\big) 
   + \mathbb{P}\big(R_{n+1}=R_{(k)}=R_{(k-1)}\big) \le \mathbb{P}\big(R_{n+1}>R_{(k-1)}\big) 
   + \epsilon_{\mathrm{tie}},
\end{equation*}
where $\epsilon_{\mathrm{tie}}$ represents the probability that the score of the $(n+1)$-th data point equals the score of any other point. Since the score distribution is continuous, $\epsilon_{\mathrm{tie}}=0$. Additionally, considering $R_{n+1}<R_{+\infty}$, we have
\begin{eqnarray*}
    & & \mathbb{P}(R_{(k-1)} < R_{n+1} < R_{+\infty}) \\
    &=& 1 - F_{P,R}(R_{(k-1)}) - [1-F_{P,R}(R_{(n)})]  \\
    &=& \alpha + [1 - \alpha - F_{P,R}(R_{(k-1)})] - \omega_{n+1} \\
    & \ge & \alpha - [F_{P,R}(R_{(k-1)}) - F_{P,R}(R_{(k)})] - \omega_{n+1} \\
    &=& \alpha - \omega_{(k)} - \omega_{n+1} \\
    & \ge & \alpha - 2 \omega_{\max}.  
\end{eqnarray*}
Consequently,
\begin{equation*}
    \mathbb{P}\big(X_{i_{\ast} j_{\ast}}\in \bar{C}(i_{\ast},j_{\ast})\big) \ge \alpha -2\omega_{\max}.
\end{equation*}
Equivalently,
\begin{equation*}
    \mathbb{P}\big(X_{i_{\ast}j_{\ast}}\in C( i_{\ast},j_{\ast})\big) \le 1-\alpha +2\omega_{\max},
\end{equation*}
which completes the proof.

\subsection{Proof of Theorem~\ref{th02}}

The proof of Theorem~\ref{th02} for full CP follows the same strategy as that of Theorem~\ref{th01}, combining ideal weighted exchangeability with a total variation bound. We therefore omit repetitive arguments and only state and prove the key auxiliary lemma required to control the upper bound of the coverage probability.

\begin{lemma}\label{lem05}
When the observation probability matrix $P$ is correctly specified and the weights are properly assigned, for any symmetric scoring function $R$ with a continuous distribution, the confidence interval $\widehat{C}$ produced by full CP for an observed location $(i_\ast,j_\ast)\mid \Omega_{\mathrm{obs}}\sim\mathrm{Unif}(\Omega_{\mathrm{obs}}^c)$ satisfies
\begin{equation*}
    \mathbb{P}\big\{X_{i_\ast j_\ast} \in \widehat{C}(i_\ast j_\ast)\big\} = \mathbb{P}\big\{X_{i_\ast j_\ast} \in \widehat{X}_{i_\ast j_\ast} \pm q_{i_\ast j_\ast}^{\ast} \cdot \widehat{\sigma}_{i_\ast,j_\ast}\big\} = \mathbb{P}\big\{R_{i_\ast j_\ast} \le q_{i_\ast j_\ast}^{\ast}\big\} \le 1 - \alpha + \omega_{\max},
\end{equation*}
where 
$$
q_{i_{\ast}j_{\ast}}^{\ast}=\operatorname{Quantile}_{1-\alpha}\Big(\sum_{k=1}^{n_{\mathrm{cal}}+1}{\omega_{i_kj_k}\delta_{R_{i_kj_k}}}\Big)
$$ 
denotes the weighted $(1-\alpha)$-quantile.
\end{lemma}

\paragraph{Proof of Lemma~\ref{lem05}.}
By weighted exchangeability, the coverage event can be written as
\begin{equation*}
X_{i_\ast,j_\ast} \in C(i_\ast,j_\ast) \iff R_{n+1}\le\operatorname{Quantile}\Big(1-\alpha; \sum_{k=1}^{n+1}\omega_k \delta_{R_k}\Big),
\end{equation*}
where $n=n_{\mathrm{cal}}$, $R_{n+1}$ denotes the score at the test point and  the scores are ordered as $R_{(1)}\le \cdots \le R_{(n+1)}$ with corresponding weights $\omega_{(1)},\ldots,\omega_{(n+1)}$. Define the weighted empirical distribution function
\begin{equation*}
     F_{P,R}(t) =\sum_{i=1}^{n+1}{\omega_i\mathbb{1}\{R_i\le t\}}.
\end{equation*}
Let
$$
R_{(k)} = \inf\{t:\widehat{F}(t)\ge 1-\alpha\}.
$$
Then the coverage event is equivalently
$$
X_{i_\ast,j_\ast}\in C(i_\ast,j_\ast) \iff R_{n+1}\le R_{(k)}.
$$
By properties of order statistics,
$$
R_{n+1}\le R_{(k)} \iff \big\{R_{n+1}<R_{(k+1)}\big\}~~\text{or}~~\big\{R_{n+1}=R_{(k)}=R_{(k+1)}\big\}.
$$
If ties occur, then $R_{n+1}=R_j$ for some $j\in[n]$. Hence,
\begin{equation*}
\mathbb{P}\big(X_{i_{\ast}j_{\ast}}\in C(i_{\ast},j_{\ast})\big) = \mathbb{P}\big(R_{n+1}<R_{(k-1)}\big) 
   + \mathbb{P}\big(R_{n+1}=R_{(k)}=R_{(k-1)}\big) \le \mathbb{P}\big(R_{n+1}<R_{(k-1)}\big) 
   + \epsilon_{\mathrm{tie}},
\end{equation*}
where $\varepsilon_{\mathrm{tie}}$ denotes the probability of ties. Since $R$ has a continuous distribution, $\varepsilon_{\mathrm{tie}}=0$. Moreover,
\begin{eqnarray*}
    & & \mathbb{P}(R_{n+1}<R_{(k+1)}) \\
    & \le & 1-\alpha+[F_{P,R}(R_{(k)}) - (1-\alpha)] \\
    & \le & 1-\alpha + [F_{P,R}(R_{(k)}) - F_{P,R}(R_{(k-1)})] \\
    &=& 1-\alpha+\omega_{(k)} \\
    & \le & 1-\alpha+\omega_{\max}.
\end{eqnarray*}
Therefore,
\begin{equation*}
  \mathbb{P}\big(X_{i_\ast j_\ast}\in C(i_\ast,j_\ast)\big) \le 1-\alpha+\omega_{\max},
\end{equation*}
which completes the proof.

\section{Additional numerical experiments}\label{appendix_ex}
 
In this section, we provide additional simulation results. We examine the effects of sample size and sparsity on the performance of the proposed method and  further investigate its behavior under misspecified observation models.

\subsection{Effects of sample size and sparsity}

Next, we investigate the effect of varying observation rates and sparsity levels $\beta$ of the sparse component $S^{\ast}$ on the performance of CP-RPCA. The following settings are considered:
\begin{itemize}
    \item \textbf{Setting 9:} Small sample size + Gaussian noise ($p=0.3$).
    \item \textbf{Setting 10:} Large sample size + Gaussian noise ($p=0.7$).
    \item \textbf{Setting 11:} Low sparsity + Gaussian noise ($\beta=0.05$).
    \item \textbf{Setting 12:} High sparsity + Gaussian noise ($\beta=0.2$).
\end{itemize}

\begin{figure}
	\centering
		\includegraphics[scale=0.45]{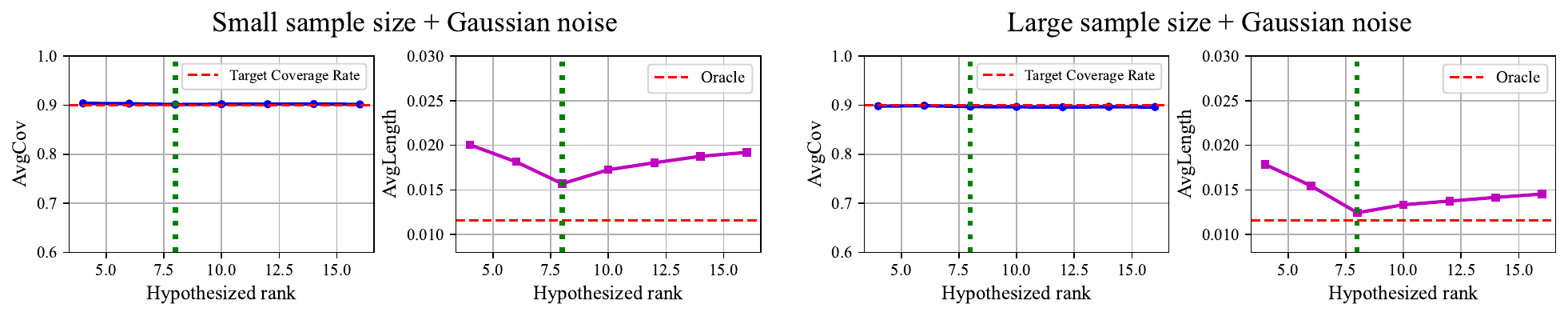}
	\caption{Comparison of effects at different sample sizes}
	\label{FIG:7}
\end{figure}

\begin{figure}
	\centering
		\includegraphics[scale=0.45]{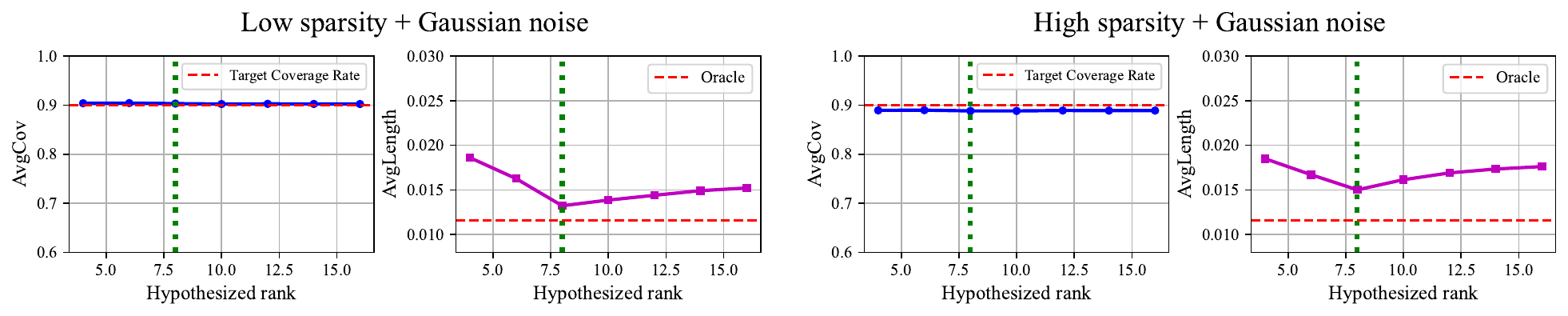}
	\caption{Comparison of effects at different sparsities}
	\label{FIG:8}
\end{figure}

Figures~\ref{FIG:7} and~\ref{FIG:8} summarize the corresponding coverage rates and average interval lengths. When the observation rate is low, the underlying RPCA estimator incurs larger reconstruction errors, leading to wider confidence intervals. Nevertheless, the empirical coverage remains close to the nominal level, indicating the stability of the conformal calibration. As the sparsity level increases, the RPCA estimation error grows only mildly, resulting in a modest increase in interval length. A slight decrease in coverage is observed in this regime, which can be attributed to occasional misclassification of contaminated entries: large noise realizations may be incorrectly identified as sparse outliers, thereby increasing the calibration error term $\xi$.

\subsection{Misspecified observation models}

We conclude by examining the robustness of CP-RPCA under misspecified observation mechanisms. Specifically, we consider the following settings:

\begin{itemize}
    \item \textbf{Setting  13:} The true missingness follows a rank-1 structure with adversarial heterogeneous noise, while a logistic missingness model is used for estimation.
    \item \textbf{Setting 14:} The true missingness follows a logistic model with adversarial heterogeneous noise, while a uniform observation model is assumed for estimation.
\end{itemize}

\begin{figure}
	\centering
		\includegraphics[scale=0.45]{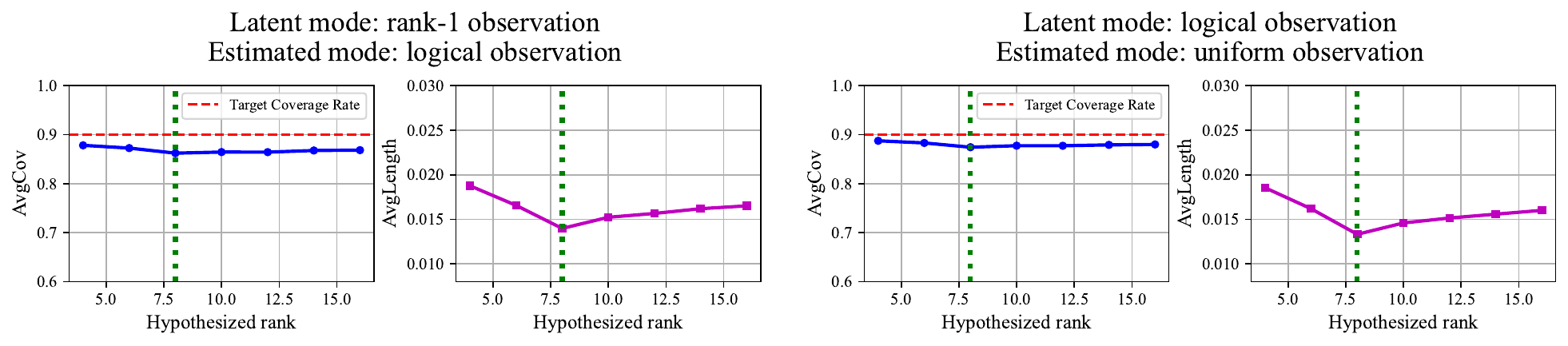}
	\caption{The performance of the model under misspecified observation patterns.}
	\label{FIG:9}
\end{figure}

Figure~\ref{FIG:9} summarizes the resulting coverage performance. As expected, both settings exhibit some degree of undercoverage due to mismatch between the true and assumed observation models, with Setting~13 showing a more pronounced degradation owing to the larger model discrepancy. Nevertheless, even in these challenging scenarios—combining observation model misspecification and adversarial heteroscedastic noise—the proposed method consistently achieves coverage rates above 85\%. These results underscore the robustness of CP-RPCA to misspecified missingness mechanisms and its ability to maintain reliable uncertainty quantification in realistic, imperfectly modeled settings.

\end{document}